\documentclass[lettersize,journal, twoside]{IEEEtran}
\usepackage{amsmath,amsfonts,amsthm,amssymb}
\usepackage{algorithmic}
\usepackage{algorithm}
\usepackage{array}
\usepackage[caption=false,font=normalsize,labelfont=sf,textfont=sf]{subfig}
\usepackage{textcomp}
\usepackage{stfloats}
\usepackage{url}
\usepackage{verbatim}
\usepackage{graphicx}
\usepackage{cite}
\usepackage{bm}
\usepackage{color}
\begin{document}

\title{Semi-supervised MIMO Detection Using\\ Cycle-consistent Generative Adversarial Network}

\author{Hongzhi Zhu,~ 
        Yongliang Guo,~ 
        Wei Xu\IEEEauthorrefmark{1},~\IEEEmembership{Senior Member,~IEEE,}~
        and Xiaohu You\IEEEauthorrefmark{1},~\IEEEmembership{Fellow,~IEEE,}
\thanks{H. Zhu is with the \textcolor{black}{National Mobile Communications Research Laboratory (NCRL), Southeast University, Nanjing, 210096, China (email: hz-zhu@seu.edu.cn).}}%
\thanks{Y. Guo is with the Purple Mountain Laboratories, Nanjing, 210096, China \textcolor{black}{(email: guoyongliang@pmlabs.com.cn).}}%
\thanks{W. Xu and X. You are with the \textcolor{black}{National Mobile Communications Research Laboratory (NCRL),} Southeast University, Nanjing, 210096, China, and the Purple Mountain Laboratories, Nanjing, 210096, China (email: \{wxu, xhyu\}@seu.edu.cn). They are also the corresponding authors of this paper.}
}

\markboth{}%
{H. Zhu \MakeLowercase{\textit{et al.}}}

\maketitle

\begin{abstract}
In this paper, a new semi-supervised deep multiple-input multiple-output (MIMO) detection approach using a cycle-consistent generative adversarial network \textcolor{black}{(CycleGAN)} is proposed for communication systems without any prior knowledge of underlying channel distributions. Specifically, we propose the \textcolor{black}{CycleGAN} detector by constructing a bidirectional loop of two modified least squares generative adversarial networks (LS-GAN). The forward LS-GAN learns to model the transmission process, while the backward LS-GAN learns to detect the received signals. By optimizing the cycle-consistency of the transmitted and received signals through this loop, the proposed method is trained online and semi-supervisedly using both the pilots and the received payload data. As such, the demand on labelled training dataset is considerably controlled, and thus the overhead is effectively reduced. Numerical results show that the proposed \textcolor{black}{CycleGAN} detector achieves better performance in terms of both bit error-rate (BER) and achievable rate than existing semi-blind deep learning (DL) detection methods as well as conventional linear detectors, especially when considering signal distortion due to the nonlinearity of power amplifiers (PA) at the transmitter.    
\end{abstract}

\begin{IEEEkeywords}
\textcolor{black}{CycleGAN}, semi-blind MIMO detection, deep learning, semi-supervised learning, nonlinearity mitigation.
\end{IEEEkeywords}

\section{Introduction}
Multiple-input multiple-output (MIMO) wireless communication system has promised significant enhancements over traditional single-antenna communication systems in terms of channel capacity and energy efficiency \cite{ref1, ref35, ref2, ref36}. While MIMO improves the performance by extending dimensions that account for time and frequency resources, it comes with new challenges of signal detection with reduced overhead to ensure reliable and efficient communications. Since signal detection is an NP-complete problem of which the optimal solution can hardly be obtained in practice, there has been growing \textcolor{black}{interest in seeking sub-optimal detection algorithms with low computational complexity} \cite{ref3}. 

\textcolor{black}{Recently,} deep learning (DL) methods have achieved great success in various fields of engineering, suggesting a new paradigm to explore data driven DL approaches for signal detection \cite{ref4}. \textcolor{black}{Existing DL-based detection methods commonly require a large number of training data, which usually leads to high overhead.} One way to overcome this \textcolor{black}{challenge} is to use off-line training. However, since off-line training typically learns a general model with prior knowledge and arbitrary assumptions, it is \textcolor{black}{usually} less robust against unknown signal \textcolor{black}{distortions} than online training. \cite{ref28}. 

In this paper, an online trained DL detection method using \textcolor{black}{a} cycle-consistent generative adversarial network \textcolor{black}{(CycleGAN)} is proposed for semi-blind MIMO detection, which outperforms typical sub-optimal detection methods with reduced overhead of pilots.

\subsection{\textcolor{black}{Classical MIMO Detection}}
Signal detector \textcolor{black}{in a MIMO communication system is} to estimate the transmitted signal from a corrupted and noisy version observed at the receiver. \textcolor{black}{The optimal detector is known as the maximum likelihood (ML) detector  which minimizes the joint probability of error} \cite{ref5, ref6, ref7}. However, for most cases, the computational complexity of \textcolor{black}{an} ML detector is too high to be practical, especially for high dimensional MIMO systems. Hence, there has been much interest in developing sub-optimal detection algorithms which balance \textcolor{black}{the} performance with acceptable computational complexity. A majority of standard researches on MIMO detection have focused on \textcolor{black}{linear detection methods, i.e.,} the matched filter (MF) detector \cite{ref8}, the zero-forcing (ZF) detector \cite{ref9}, and the linear minimum mean-squared error (LMMSE) \mbox{detector \cite{ref9, ref10, ref11}}. While such linear methods indicate relatively low computational complexity, they typically require accurate channel estimation and known \textcolor{black}{statistics of the noise. For communications with nonlinear signal distortion and mathematically intractable channel models, the performance of linear detection methods encounters considerable degradation \cite{ref12}.}

\subsection{\textcolor{black}{DL-based MIMO Detection}}
While classic detection algorithms are typically based on detection theory which essentially depends on the assumption of a prior probabilistic model of the transmission process, DL-based detection methods require no prior knowledge of the underlying channel model and \textcolor{black}{they} learn the transmission process by minimizing the statistic loss function with known samples of signals. To be specific, \textcolor{black}{ the signal detection problem in DL} is equivalent to a classification problem which outputs a soft decision for a given noisy and corrupted version of the desired signal.

DL models can learn directly from data \textcolor{black}{samples. Thus, they have a promising potential for conducting MIMO detection} under complex scenarios which can not be explicitly modeled by mathematical \textcolor{black}{formulations} and where the channel models are unknown \cite{ref34}. In \cite{ref13}, a deep MIMO detection method, \textcolor{black}{namely the Detection Network (DetNet)}, based on deep neural network (DNN) was proposed as an early attempt. While DetNet shows noticeable robustness with low computational complexity under ill-conditioned channels, it still requires instantaneous channel state information (CSI) and its performance degrades if \textcolor{black}{the} CSI estimate is inaccurate. In \cite{ref14}, a modified DNN-based detector was proposed \textcolor{black}{which outperforms not only DetNet but also a similarly-structured detector based on convolutional neural networks (CNN). However, all these DL-based detectors require accurate CSI estimation, which imposes extra effort \textcolor{black}{in practice}.}

To avoid performance degradation caused by inaccurate CSI estimation, more complex DL models have been explored for blind or semi-blind signal detection. In \cite{ref15}, a blind channel equalizer based on variational autoencoder (VAE), \textcolor{black}{referred to as} VAEBCE, was proposed, which achieves BER performance close to \textcolor{black}{a} non-blind adaptive LMMSE equalizer. \textcolor{black}{The} VAEBCE models the transmission process without CSI estimation but it requires a large amount of training data which means it has to work under fixed channel to obtain \textcolor{black}{enough data samples. A modified turbo equalizer, namely the turbo VAEE, was later devised in \cite{ref16} which outperforms the turbo Expectation-Maximization (EM) algorithm under various invariant channels within a long coherence time.} 

\textcolor{black}{In most wireless communication systems}, the channel dynamically changes and it is impractical to obtain a large scale of training dataset under a specific \textcolor{black}{invariant} channel before detection. Therefore, off-line trained DL methods have been \textcolor{black}{investigated for practical uses. Such methods can be trained off-line as a general model using simulation data or experimental collections if the channel stably maintains the features indicated by the training data, e.g., the signal distortion and channel
variation timescale.} In \cite{ref17}, a DL-based joint channel estimation and signal detection algorithm for \textcolor{black}{an orthogonal frequency division multiplexing (OFDM)} system was proposed, which achieved superior performance \textcolor{black}{compared with conventional methods. This method used a channel estimation network (CENet) for channel estimation and a channel conditioned recovery network (CCRNet)} for signal detection. However, the CENet and CCRNet indicate \textcolor{black}{an} urgent requirement for a large training dataset and their performance degrades even for high \textcolor{black}{signal-to-noise ratios (SNR) if the signals} under the operating SNR are not included in the training dataset. In \cite{ref18}, a compressed sensing (CS) approach using wasserstein generative adversarial network (WGAN) was explored to conduct high dimensional channel estimation, which achieved a significant performance gain over typical techniques for sparse signal recovery such as \textcolor{black}{the} orthogonal matching pursuit (OMP) algorithm and \textcolor{black}{the} approximate message passing (AMP) algorithm. \textcolor{black}{However, these off-line trained methods can encounter noticeable performance degradation if the varying channel distribution is no longer consistent with the training dataset.}

To better conduct signal detection under channels which are completely unknown \textcolor{black}{and} difficult to be derived analytically, one possible way is to explore online-trained methods which can adapt themselves to \textcolor{black}{varying} channels with reduced overhead.        

\begin{figure*}[!t]
\centering
\includegraphics[width=5 in]{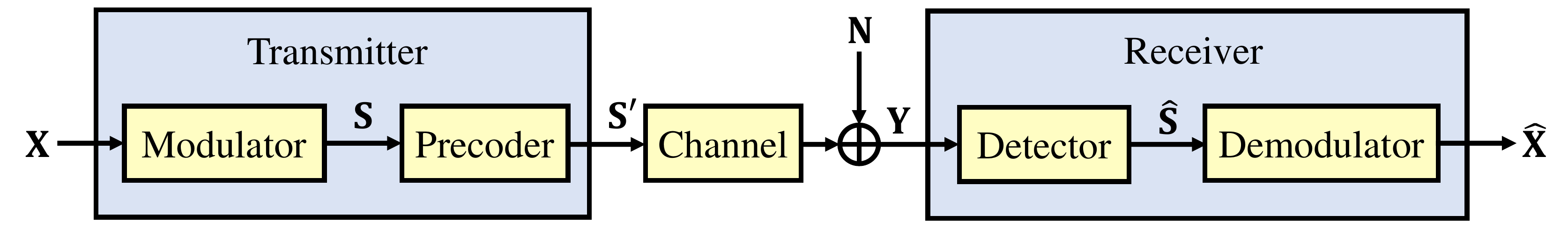}
\caption{\textcolor{black}{MIMO downlink communication system model. The source bit sequence $\mathbf{X}$ is modulated into $\mathbf{S}$ and precoded into $\mathbf{S^{\prime}}$ at the transmitter. $\mathbf{S^{\prime}}$ is then passed through the channel and is received as $\mathbf{Y}$. At the receiver an estimation of $\mathbf{S}$ is obtained as $\mathbf{\hat{S}}$ by a detector basing on $\mathbf{Y}$ and is then demodulated into $\mathbf{\hat{X}}$ as a recovery of the source bit sequence. }}
\label{fig_system}
\end{figure*}


\begin{figure}[!t]
\centering
\includegraphics[width=3.4in]{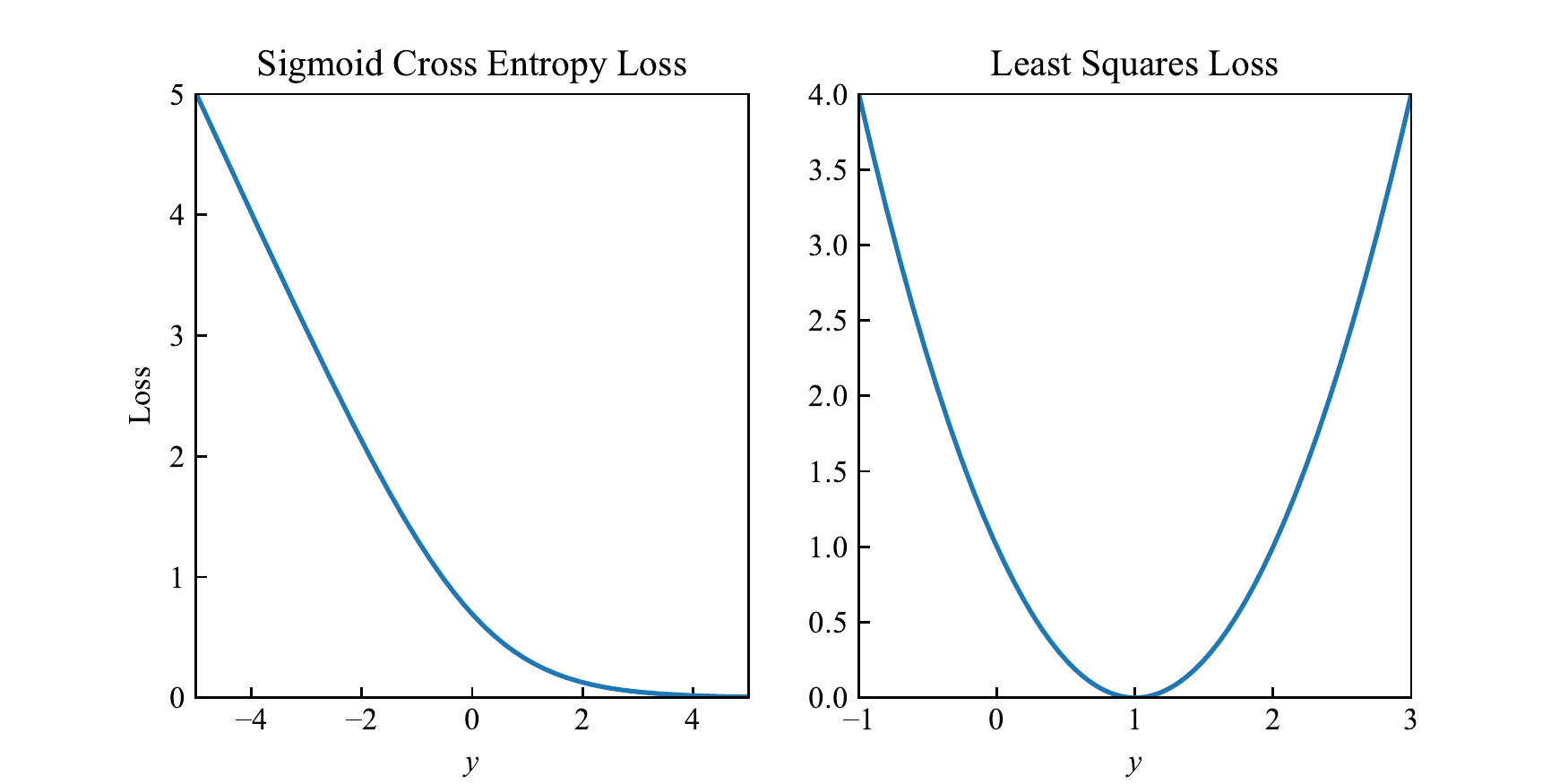}
\caption{\textcolor{black}{Curves of (a) the sigmoid cross entropy loss and (b) the least squares loss. For the misidentified generated samples on the positive side of the discriminator, the least squares loss retains the error while the sigmoid cross entropy loss remains close-to-zero for ${\bm y} \ge 2.5$.}}
\label{fig_loss}
\end{figure}

\subsection{Contributions}
Since standard detection methods may face difficulty when the underlying channel model is unknown and \textcolor{black}{hard to be tracked with limited overhead, the best approach to designing detection algorithms remains to be explored.} In this paper, we propose a novel semi-blind MIMO signal detection method using \textcolor{black}{CycleGAN}, which is trained online using \textcolor{black}{both the} pilots and received payload data to keep tracking the \textcolor{black}{varying channel.} The major contributions of our work are summarized as follows:
\begin{itemize}
\item{The \textcolor{black}{CycleGAN} detector requires neither prior knowledge of the channel nor prior simulation data for off-line training. It performs well even \textcolor{black}{under unknown channel effects such as nonlinear distortion due to imperfect power amplifiers (PA) at the transmitter.}}
\item{Different from typical \textcolor{black}{DL-based detection methods which require large overhead for training data, the number of pilots needed to train the proposed \textcolor{black}{CycleGAN} detector is no more than the number of transmit antennas as the popular setup in practice \cite{ref33}. Moreover, the \textcolor{black}{CycleGAN} detector does not require explicit channel estimation, which further reduces the practical effort.}}
\item{We propose a bidirectional network loop for the \textcolor{black}{CycleGAN} detector to check the cycle-consistency of signals, where the received payload data are also \textcolor{black}{used to train the model in an unsupervised phase.} The detection accuracy is therefore improved and the overfitting caused by the lack of training data is efficiently avoided.}
\end{itemize}

Numerical results show that \textcolor{black}{the proposed \textcolor{black}{CycleGAN} detector achieves better BER and achievable rate than other benchmarks, especially for scenarios with nonlinear PA distortion. Experiments also illustrate that the introduction of semi-supervised learning using both the pilots and the received payload data brings noticeable improvement on the pilot reduction than typical pilot-aided DL-based detection methods.} 

\subsection{Paper Outline}
The rest of this paper is organized as follows. In Section \uppercase\expandafter{\romannumeral2}, we present the system model and \textcolor{black}{describe} the MIMO detection problem. In Section \uppercase\expandafter{\romannumeral3}, the mechanism of the proposed \textcolor{black}{CycleGAN} detector is introduced with the training strategy. Implementation details of the \textcolor{black}{CycleGAN} detector is elaborated in Section \uppercase\expandafter{\romannumeral4}, including the pre-processing of dataset and \textcolor{black}{empirical training setups.}  \textcolor{black}{The detection performance of the proposed \textcolor{black}{CycleGAN} detector is evaluated with comparisons in Section \uppercase\expandafter{\romannumeral5}.} Conclusions are drawn in Section \uppercase\expandafter{\romannumeral6}.

\section{Prerequisites}
In this section, we introduce the MIMO communication system model and present the problem formulation of MIMO detection.

\subsection{System Model}
We consider the downlink of a MIMO system as shown in Fig. \ref{fig_system} with $N_{\rm s}$ transmit antennas and $N_{\rm r}$ receive antennas. In this figure, $\mathbf{X}$ is the source bit sequence, $\mathbf{S}$ is the modulated signal, and $\mathbf{S^{\prime}}$ is the precoded signal for transmission. The precoding process can be formulated as
\begin{equation}
\label{mimo_model0}
\mathbf{S^{\prime}} = \mathbf{F}\mathbf{S} \text{,}
\end{equation}
where $\mathbf{F}$ is the precoding matrix. At the receiver side, \textcolor{black}{the addictive noise is denoted by $\mathbf{N}$, and the received signal is denoted by $\mathbf{Y}$.}

The standard linear input-output \textcolor{black}{relationship of a} MIMO communication system is given by
\begin{equation}
\label{mimo_model1}
\mathbf{Y} = \mathbf{H}\mathbf{F}\mathbf{S} + \mathbf{N} \text{,}
\end{equation}
where $\mathbf{H} \in \mathbb{C}^{N_{\rm r} \times N_{\rm s}}$ is the MIMO channel matrix. 

However, considering nonlinear distortion in the system \textcolor{black}{due to the peak-to-average power ratio reduction in OFDM systems} \cite{ref19}, low precision \textcolor{black}{analog-to-digital converter (ADC) quantization} \cite{ref20}, \textcolor{black}{and nonlinearity of PA} \cite{ref21}, \textcolor{black}{in more general scenarios, the system modeled in} (\ref{mimo_model1}) is reformulated by
\begin{equation}
\label{mimo_model2}
\mathbf{Y} = f_{\rm{H}}(\mathbf{S}) + \mathbf{N} \text{,}
\end{equation}
where $f_{\rm{H}}(\cdot)$ is a nonlinear function which models the entire transmission process including the precoding, physical channel, and nonlinear distortion.

The received signal $\mathbf{Y}$ is always noisy and corrupted by the channel. Since the channel is unknown and changes in time, pilots need to be periodically inserted in the transmitted signals for the receiver to estimate the channel. The goal of the detector is to reconstruct the payload $\mathbf{S}$ from the observed $\mathbf{Y}$ and the pilots. After detection, the estimated $\mathbf{\hat{S}}$ is passed to the demodulator to recover the source bit sequence $\mathbf{X}$, denoted \textcolor{black}{by} $\mathbf{\hat{X}}$.

\subsection{MIMO Detection}
In a typical MIMO system, \textcolor{black}{signal detection is conducted after each block} which contains a \textcolor{black}{pilot-training} period and a \textcolor{black}{data-transmission} period. We assume a block-fading scenario with time division duplex (TDD) mode and a coherence time transmitting $K = P + D$ symbols, \textcolor{black}{where the first $P$ symbols are the pilot sequence $\mathbf{S_{\rm P}}$ and the following $D$ symbols are the payload data $\mathbf{S_{\rm D}}$.} Accordingly, the received signal in (\ref{mimo_model2}) \textcolor{black}{is} rewritten as 
\begin{equation}
\label{mimo_model3}
[\mathbf{Y}_{\rm{P}},\mathbf{Y}_{\rm{D}}] = f_{\rm{H}}([\mathbf{S}_{\rm{P}}, \mathbf{S}_{\rm{D}}]) + \mathbf{N} \text{,}
\end{equation}
where $\mathbf{Y}_{\rm{P}}$ and $\mathbf{Y}_{\rm{D}}$ are \textcolor{black}{the received pilots and received payload data, respectively.} In the \textcolor{black}{pilot-training} period of $P$ symbols, both $\mathbf{Y}_{\rm{P}}$ and $\mathbf{S}_{\rm{P}}$ are known, which allows \textcolor{black}{supervised learning to model $f_{\rm{H}}(\cdot)$.} In the subsequent \textcolor{black}{data-transmission} period of $D$ symbols, standard pilot-aided detection methods detect the payload $\mathbf{S}_{\rm{D}}$ directly with $\mathbf{Y}_{\rm{D}}$ \textcolor{black}{using the model trained by pilots.} However, for a relatively complex $f_{\rm{H}}(\cdot)$, \textcolor{black}{the model trained by pilots has a high risk of overfitting due to the limitation of pilots, and the performance therefore encounters degradation. For the proposed \textcolor{black}{CycleGAN} detector, we coupled $\mathbf{S}_{\rm{P}}$, $\mathbf{Y}_{\rm{P}}$, and $\mathbf{Y}_{\rm{D}}$ together for an extra semi-supervised training phase so that the model is further updated to fit the payload data. After the semi-supervised training phase, a better estimation of $\mathbf{S}_{\rm{D}}$ is obtained by the latest model, and thus the performance is effectively improved.}
%


\section{CycleGAN Detector}
In this section, we formulate the mechanism of the \textcolor{black}{CycleGAN} detector, including the architecture of \textcolor{black}{CycleGAN} and the training strategy.

\textcolor{black}{To conduct signal detection, it is necessary to model the effective channel, i.e., the relationship between the transmitted signal and the received signal. Standard channel models and detection algorithms are based on explicit mathematical tools, which can face difficulty on covering unknown channel effects, thus suggesting exploration of DL models. During communications, the detector has to generate new signals which are not learned from the limited number of pilots. Thus, generative models are rather preferred than classic DL models such as basic DNN and CNN which are commonly used in classification tasks.}

\textcolor{black}{In Section \ref{ls-gan}, we propose a modified LS-GAN as a basic model of the effective channel. Then in Section \ref{cycleGAN}, we set up a bidirectional loop of two modified LS-GANs as a \textcolor{black}{CycleGAN} to enable unsupervised learning using the received payload data. In Section \ref{train_scheme}, we elaborate the entire semi-supervised training strategy of the proposed \textcolor{black}{CycleGAN} detector. Computational complexity analysis of the proposed \textcolor{black}{CycleGAN} is presented in Section \ref{complexity}.}

\begin{table*}[!t]
\renewcommand\arraystretch{1.5}
\caption{Realizations of F-GAN\label{table_fGAN}}
\centering
\begin{tabular}{ccc}
\hline
Name & $D_{\rm f}(P||Q)$ & Generator $f(y)$  \\
\hline
Kullback-Leibler & $\int p(x){\rm log}\frac{p(x)}{q(x)}{\rm d}x$ & $y {\rm log}y$\\
Reverse KL & $\int q(x){\rm log}\frac{q(x)}{p(x)}{\rm d}x$ & $-{\rm log}y$ \\
Pearson $\chi ^ 2$ & $\int \frac{(q(x)-p(x))^2}{p(x)}{\rm d}x$ & $(y-1)^2$  \\
Squared Hellinger & $\int \left(\sqrt{p(x)}-\sqrt{q(x)}\right)^2{\rm d}x$  & $(\sqrt{y}-1)^2$\\
Jensen-Shannon  & $\frac{1}{2}\int [p(x){\rm log}\frac{2p(x)}{p(x)+q(x)} + q(x){\rm log} \frac{2q(x)}{p(x)+q(x)}]{\rm d}x$  &  $-(y+1){\rm log}\frac{1+y}{2} + y {\rm log}y$\\
GAN & $\int [p(x){\rm log}\frac{2p(x)}{p(x)+q(x)} + q(x){\rm log}\frac{2q(x)}{p(x)+q(x)}]{\rm d}x - {\rm log}(4)$ & $y{\rm log}y - (y+1){\rm log}(y+1)$ \\
\hline
\end{tabular}
\end{table*}

\begin{figure*}[!t]
\centering
\includegraphics[width=5 in]{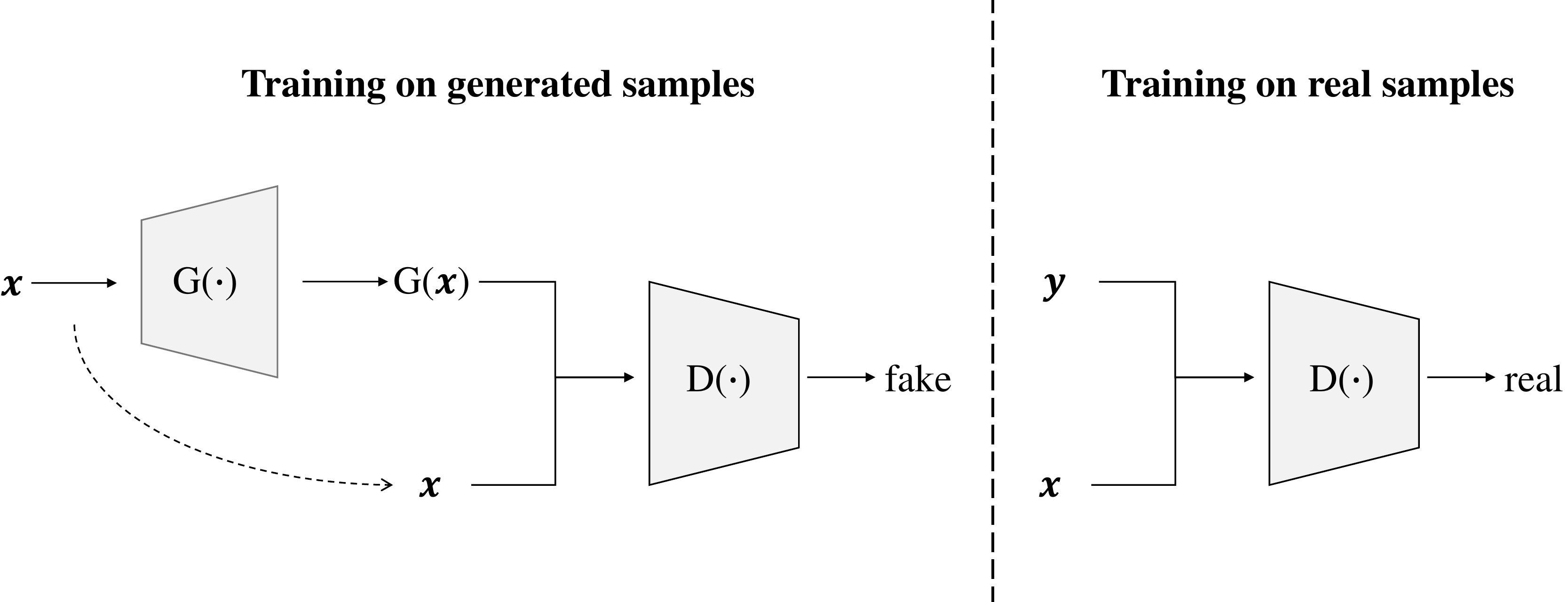}
\caption{\textcolor{black}{Architecture of the proposed LS-GAN, consisting of a generator $\rm{G}(\cdot)$ and a discriminator $\rm{D}(\cdot)$. Both $\rm{G}(\cdot)$ and $\rm{D}(\cdot)$ are implemented by deep neural networks (DNN) and are trained under an adversarial training strategy, \textcolor{black}{i.e.,} the discriminator learns to distinguish the generated samples while the generator learns to create better samples that do not be distinguished.}}
\label{fig_lsgan}
\end{figure*}

\subsection{LS-GAN\label{ls-gan}}

\textcolor{black}{LS-GAN is an extension of generative adversarial network (GAN), which uses the least squares distance as the training loss instead of sigmoid cross entropy. Classical LS-GAN has been proved as a more stable unsupervised learning method with better performance than typical generative models \cite{ref22}.} 

\textcolor{black}{Actually there have been a considerable amount of GAN extensions and each has its own merits. In \cite{ref32}, a series of GAN extensions are induced as general f-GAN, and the objective loss functions of those extensions are induced as f-divergence. Table \ref{table_fGAN} lists various realizations of the f-GAN, where $D_{\rm f}(P||Q)$ is the f-divergence for optimizing in theory, $f(y)$ is the objective function of the generator, and $y$ is the output of the discriminator. As shown in \mbox{Table \ref{table_fGAN}}, the \mbox{Pearson $\chi ^ 2$} and Squared Hellinger divergence metrics aim at reducing the least squares loss and its extensions, respectively, while the other realizations use extensions of a sigmoid cross entropy loss as objective functions. Note that the least squares loss used by LS-GAN is equivalent to the Pearson $\chi ^ 2$ \mbox{divergence \cite{ref22}}.}

\textcolor{black}{One major advantage of LS-GAN over other realizations of f-GAN is that it has a quite lower risk of encountering the vanishing gradient problem \cite{ref22}. Fig. \ref{fig_loss} shows the curves of a sigmoid cross entropy loss and a least squares loss, respectively, where $y$ is the output of the discriminator. In the early stage of training, the discriminator is too weak to separate generated samples, i.e., there can be a considerable amount of generated samples on the positive side of the discriminator but still far away from the real data. For such samples the sigmoid cross entropy loss yields small close-to-zero errors as in Fig. \ref{fig_loss}(a), which brings high risk of vanishing gradient problem. In contrast, the least squares loss can yield noticeable errors as shown in Fig. \ref{fig_loss}(b), and it has only one flat point at zero. In this way, it is less likely to encounter the vanishing gradient problem by LS-GAN.}

\textcolor{black}{In massive MIMO systems, the high dimension of signals and inevitable existence of noise make it quite difficult for the discriminator to separate generated samples. The positive output of the discriminator to those generated samples causes insignificant errors through a sigmoid cross entropy loss. Such insignificant errors can lead to inadequate guidance for the generator or even result in vanishing gradient problem. Actually, in our experiments the classical GAN can hardly complete a training due to the high possibility of encountering the vanishing gradient problem, and this risk of failure becomes considerably higher for a large MIMO of size $64 \times 8$ than $2 \times 2$. In contrast, the least squares loss can effectively avoid the vanishing gradient problem since it yields much more significant errors on the positive output of the discriminator for those misidentified generated samples.} 

\textcolor{black}{Moreover, the least squares loss can yield more noticeable error for samples far from the real data than a sigmoid cross entropy loss since the distance between the generated samples and the real samples is directly squared rather than transformed into probability to calculate entropy. This may cause degradation in generative diversity especially for image generation cases, i.e., the generator tends not to create various samples for the same conditional input \cite{ref22}. In MIMO detection, however, the pairwise relationship between the transmitted and received signals is deterministic under a given channel without noise. That is to say, the model should focus on the exact conditional distribution between signals and it does not need to create various artificial results.}

\textcolor{black}{Thus, we consider that the major features of LS-GAN are more suitable than other GAN methods for solving the MIMO detection problem.}

\textcolor{black}{MIMO detection can be elaborated as a conditional problem which requires to model the mapping relationship between the transmitted signal and the received signal. Since the two signals are coupled as a pair during the transmission, we adapted the classical LS-GAN for MIMO detection with some modifications to both the generator and the discriminator. Furthermore, we propose a \textcolor{black}{CycleGAN} for MIMO detection by setting up a bidirectional loop of two LS-GANs with opposite input and output, which forms a semi-supervised learning method. For this sake of explanation, we denote the desired output by ${\bm y}$ and the input by ${\bm x}$ for each LS-GAN. That is to say, for the forward LS-GAN, ${\bm x}$ represents the transmitted signal and ${\bm y}$ represents the received signal, while for the backward LS-GAN the representations are reversed. The following are the modifications in the generator and the discriminator of each LS-GAN in the proposed \textcolor{black}{CycleGAN}.} 
\subsubsection{\textcolor{black}{Generator}} 
Classical conditional extensions of LS-GAN and GAN for mapping of ${\bm x} \xrightarrow{} {\bm y}$ typically add an extra random noise $\bm{z}$ to the input of the generator to enable the generation of variable samples\cite{ref30, ref31}. Taking image style translation as an example, a conditional GAN is able to transform the same sketch into many oil paintings in various colors using different $\bm{z}$. The introduction of $\bm{z}$ requires unsupervised learning since \textcolor{black}{classical models} actually learn the mapping of $({\bm x}, {\bm z}) \xrightarrow{} {\bm y}$. However, in MIMO detection, the transmitted signal and the received signal inherently corresponds to each other via a noisy channel. Thus, we removed the component of $\bm{z}$ and only input $\bm{x}$ to generate $\bm{y}$ for each of the LS-GANs. Then the proposed LS-GANs are able to model the direct mapping of ${\bm x} \xrightarrow{} {\bm y}$ by supervised learning using the coupled data pairs, i.e., $[{\bm x}, {\bm y}]$.
\subsubsection{\textcolor{black}{Discriminator}} 
\textcolor{black}{In order to focus on the pairwise relationship between the transmitted signal and the received signal, we concatenate $\bm{x}$ and $\bm{y}$ as an entire input of the discriminator for each LS-GAN. In this way the discriminator judges whether the entire data pair $[{\bm x}, {\bm y}]$ is \textcolor{black}{from the} real data, and thus the signals from two sides are learned as pairs.}

\textcolor{black}{The entire architecture of the proposed LS-GAN is shown in Fig. \ref{fig_lsgan}.} An LS-GAN consists of a generator $\rm{G}(\cdot)$ and a discriminator $\rm{D}(\cdot)$, each of which is implemented by a DNN. For each single vector of input data $\bm{x}$, $\rm{G}(\cdot)$ outputs an estimated sample $\bm{y}^{\prime}$, as
\begin{equation}
\label{LS-1}
\bm{y}^{\prime} = {\rm G}(\bm{x}) \text{.}
\end{equation}
Then in order to make $\bm{y}^{\prime}$ an accurate estimate of the corresponding real output $\bm{y}$, we \textcolor{black}{couple the generated data pair $[\bm{x}$, $\bm{y}^{\prime}]$ and the real data pair $[\bm{x}$, $\bm{y}]$ together to train $\rm{D}(\cdot)$ for global discrimination \cite{ref23}. The output of $\rm{D}(\cdot)$ is a scalar which indicates whether the input data pair is from the real data.} 

\textcolor{black}{Now we give a detailed formulation of the LS-GAN model. Firstly, we focus on the generator $\rm{G}(\cdot)$, which is realized by a DNN parameterized by $\bm{\theta}$. The input data of $\rm{G}(\cdot)$ is considered as a sample data $\bm{x}$ obeying an input-distribution $P_{X}(\bm{x})$, and the output of $\rm{G}(\cdot)$ is considered as another sample data $\bm{y}^{\prime}$ obeying the generate-distribution $P_{\rm{G}}(\bm{y} \vert \bm{x};\bm{\theta})$. Assuming that the real output-distribution is $P_{Y \vert X}(\bm{y} \vert \bm{x})$, the purpose of $\rm{G}(\cdot)$ is equivalent to training $P_{\rm{G}}(\bm{y} \vert \bm{x};\bm{\theta})$ to approximate $P_{Y \vert X}(\bm{y} \vert \bm{x})$ by calculating the best parameter $\bm{\theta ^ {\ast}}$ by
\begin{equation}
\label{GAN_func1}
\bm{\theta^{\ast}} = \arg\mathop{\min}\limits_{\bm{\theta}}{\mathcal{L}_{0}}(P_{Y \vert X}(\bm{y} \vert \bm{x}) \Vert P_{\rm{G}}(\bm{y} \vert \bm{x};\bm{\theta})) \text{,}
\end{equation}
where ${\mathcal{L}_{0}}(P_{Y \vert X}(\bm{y} \vert \bm{x}) \Vert P_{\rm{G}}(\bm{y} \vert \bm{x};\bm{\theta}))$ is the least squares distance between $P_{Y \vert X}(\bm{y} \vert \bm{x})$ and $P_{\rm{G}}(\bm{y} \vert \bm{x};\bm{\theta})$.}     

\textcolor{black}{Then we elaborate the mechanism of the discriminator $\rm{D}(\cdot)$, which is a DNN classifier parameterized by $\bm{\phi}$. The training dataset of $\rm{D}(\cdot)$ contains real data pairs $[\bm{x}, \bm{y}]$ and generated data pairs $[\bm{x}, \bm{y}^{\prime}]$. The purpose of $\rm{D}(\cdot)$ is equivalent to solving the binary classification problem, where the distance between a real data pair and a generated data pair is defined by the squared error as
\begin{equation}
\begin{split}
\label{GAN_func}
{\mathcal{L}_{1}}({\rm G},{\rm D}) = &\mathbb{E}_{\bm{x} \sim P_{X}, \bm{y} \sim P_{Y \vert X}}[({\rm D}([\bm{x}, \bm{y}]) - 1)^{2}\\ 
&+ ({\rm D}([\bm{x}, {\rm G}(\bm{x})]) + 1)^{2}] \text{,}
\end{split}
\end{equation}
which is set as the objective function for optimizing ${\rm D}(\cdot)$. Taking the derivative of ${\mathcal{L}_{1}}({\rm G}, {\rm D})$ and forcing it to zero, we obtain the optimal ${\rm D}(\cdot)$ minimizing ${\mathcal{L}_{1}}({\rm G}, {\rm D})$ as 
\begin{equation}
\begin{split}
\label{GAN_func5}
{\rm D}^{\ast}([\bm{x}, \bm{y}]) = \frac{p_{Y \vert X}(\bm{y} \vert \bm{x}) - p_{{\rm G}}(\bm{y} \vert \bm{x})}{p_{Y \vert X}(\bm{y} \vert \bm{x}) + p_{{\rm G}}(\bm{y} \vert \bm{x})} \text{,}
\end{split}
\end{equation}
where $p_{Y \vert X}$ and $p_{{\rm G}}$ are the probability density functions (PDF) of $P_{Y \vert X}$ and $P_{\rm{G}}$ respectively.}   

\begin{figure*}[t]
\centering
\subfloat[The proposed \textcolor{black}{CycleGAN}.]{
    \centering
    \begin{minipage}[t]{0.3\linewidth}
        \centering       
        \includegraphics[width=\linewidth]{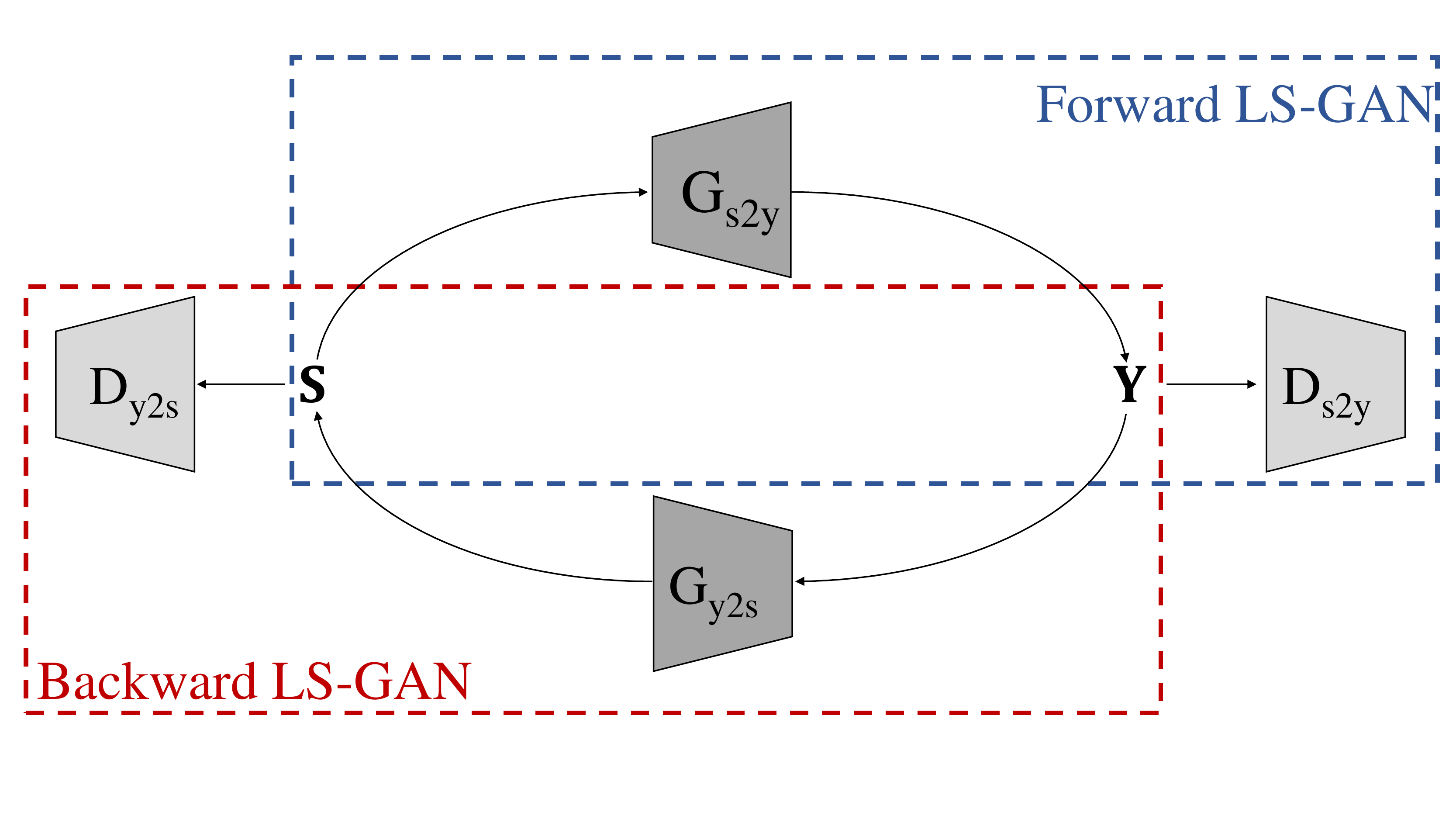}
        \end{minipage}
}%
\subfloat[Forward consistency loss.]{
    \centering
    \begin{minipage}[t]{0.3\linewidth}
        \centering       
        \includegraphics[width=\linewidth]{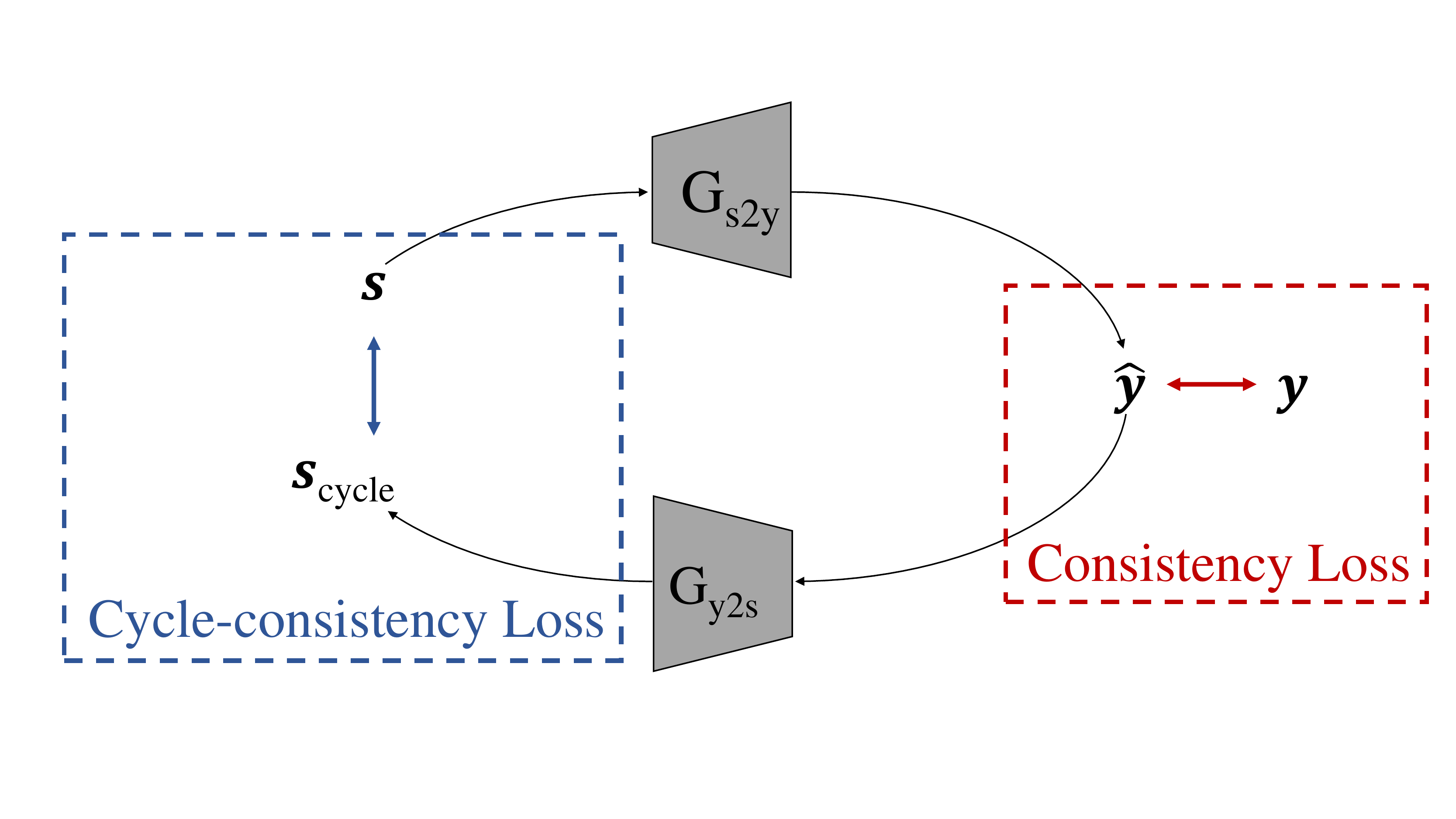}
    \end{minipage}
}%
\subfloat[Backward consistency loss.]{
    \centering
    \begin{minipage}[t]{0.3\linewidth}
        \centering       
        \includegraphics[width=\linewidth]{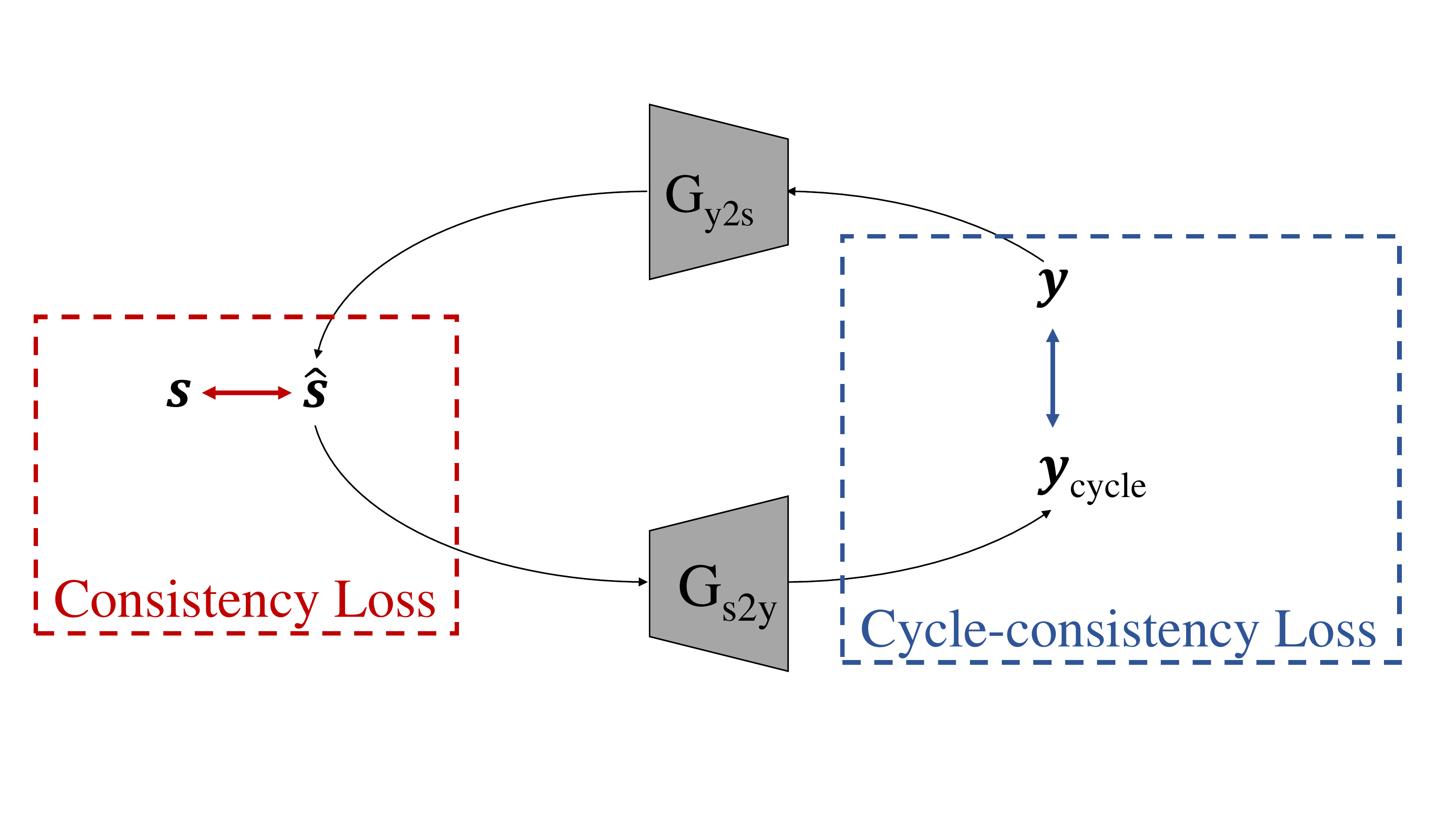}
    \end{minipage}
}%

\caption{\textcolor{black}{(a) Architecture of the proposed \textcolor{black}{CycleGAN}. Two LS-GANs are coupled into a bidirectional loop so that the input data could be sent to another side and back again for checking the cycle-consistency. The \textcolor{black}{CycleGAN} introduces a forward loss as (b) to check the consistency in the path of $\mathbf{S} \xrightarrow{} \mathbf{Y} \xrightarrow{} \mathbf{S}$ and a backward loss as (c) to check the consistency in the path of $\mathbf{Y} \xrightarrow{} \mathbf{S} \xrightarrow{} \mathbf{Y}$.}}
\label{fig_cycleGAN}
\end{figure*}

\textcolor{black}{Then from the perspective of ${\rm G}(\cdot)$, since $P_{Y \vert X}(\bm{y} \vert \bm{x})$ is unknown, we represent the least squares distance in (\ref{GAN_func1}) by the output of ${\rm D}(\cdot)$ as
\begin{equation}
\label{GAN_func2}
{\mathcal{L}_{0}}(P_{Y \vert X}(\bm{y} \vert \bm{x}) \Vert P_{\text{G}}(\bm{y} \vert \bm{x};\bm{\theta})) \triangleq \mathbb{E}_{\bm{x} \sim P_{X}}[{\rm D}([\bm{x}, {\rm G}(\bm{x})])^{2}] \text{,}
\end{equation}
which is set as the objective function for optimizing ${\rm G}(\cdot)$. So far we have transformed the problem of solving unknown $P_{Y \vert X}(\bm{y} \vert \bm{x})$ into a problem of minimizing the squared expectation of ${\rm D}([\bm{x}, {\rm G}(\bm{x})])$, which yields the optimal ${\rm G}^{\ast}(\bm{x})$ with the optimal $\bm{\theta}^{\ast}$ given by}
\begin{equation}
\begin{split}
\label{GAN_func3}
\bm{\theta}^{\ast} = \arg\mathop{\min}\limits_{{\bm{\theta}}}\mathbb{E}_{\bm{x} \sim P_{X}}[{\rm D}([\bm{x}, {\rm G}(\bm{x};{\bm{\theta}})])^{2}] \text{.}
\end{split}
\end{equation}

\textcolor{black}{During the training, the discriminator ${\rm D}(\cdot)$ aims at minimizing (\ref{GAN_func}) with fixed ${\rm G}(\cdot)$ so that it could be able to identify whether the input data is from the real data, while the generator ${\rm G}(\cdot)$ aims at minimizing (\ref{GAN_func2}) with fixed ${\rm D}(\cdot)$ to avoid the generated data from negative judgement.}  

\begin{figure*}[!t]
\centering
\includegraphics[width=4.5in]{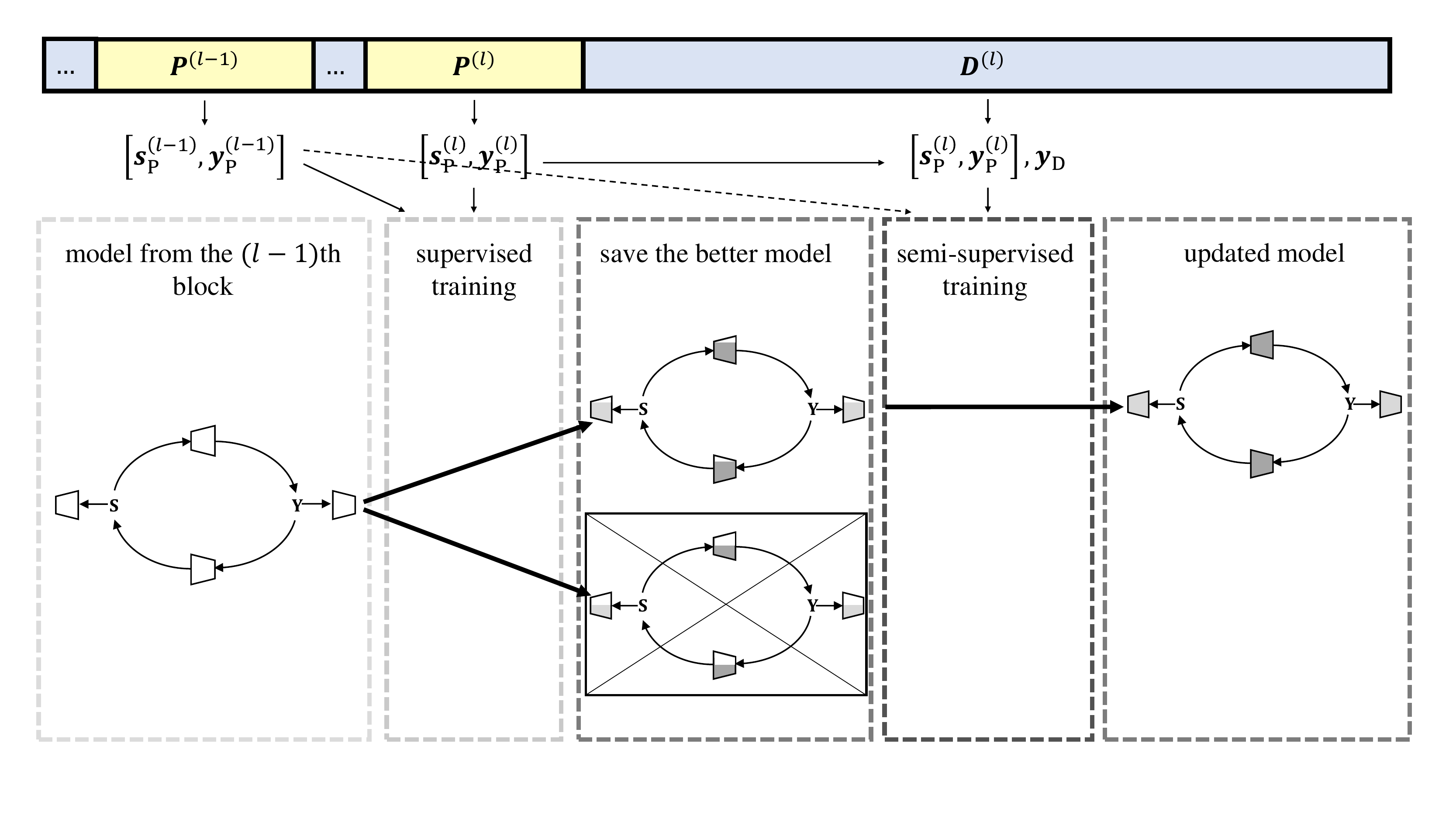}
\caption{\textcolor{black}{Training strategy of the model for the $l$th block of $K = P + D$ time slots. In the pilot-training period the pilots $[\bm{s}^{(l)}_{\rm P}, \bm{y}^{(l)}_{\rm P}]$ are used for a supervised training phase. Then in the data-transmission period both the pilots $[\bm{s}^{(l)}_{\rm P}, \bm{y}^{(l)}_{\rm P}]$ and received payload data $\bm{y}_{\rm D}$ are used for a semi-supervised training phase. The previous pilots $[\bm{s}^{(l-1)}_{\rm P}, \bm{y}^{(l-1)}_{\rm P}]$ are also used for training if they could make the model a progress.}}
\label{fig_train}
\end{figure*}

\subsection{CycleGAN \label{cycleGAN}}
\textcolor{black}{CycleGAN was originally proposed as an image-to-image translation model using a bidirectional loop of LS-GANs to realize image style-conversion \cite{ref24}.} Since signal detection could be equivalent to a signal-to-signal translation, the core mechanism of \textcolor{black}{CycleGAN} can bring some guidance for designing the detector.

We \textcolor{black}{construct} two proposed LS-GANs as a bidirectional loop to conduct signal detection, of which the forward LS-GAN models the effective channel and the backward LS-GAN models the detector. The two LS-GANs are set with similar network structures, but trained with converse input and output. The generator and the discriminator of the forward LS-GAN are denoted by $\rm{G}_{s2y}(\cdot)$ and $\rm{D}_{s2y}(\cdot)$, respectively, while those of the backward LS-GAN are denoted by $\rm{G}_{y2s}(\cdot)$ and $\rm{D}_{y2s}(\cdot)$, respectively.

\textcolor{black}{Since real data is a necessity for the proposed LS-GANs, we first train them supervisedly by pilots in the pilot-training period. However, the number of pilots is always limited in practice. Thus, we introduce the architecture of \textcolor{black}{CycleGAN} to realize semi-supervised learning using both the pilots and the received payload data.} 

The proposed \textcolor{black}{CycleGAN} model coupled the two LS-GANs mentioned in Section \ref{ls-gan} into a bidirectional loop as illustrated in Fig. \ref{fig_cycleGAN}(a), by which the input data could be sent to another side and back again for checking the cycle-consistency. \textcolor{black}{In this way, during the data-transmission period, the model continues to be updated using both the pilots and the received payload data, even though the real payload data pairs are unknown.}  

For a single data pair $[\bm{s}, \bm{y}]$, besides the basic loss function of  \textcolor{black}{the proposed LS-GAN in (\ref{GAN_func}) and (\ref{GAN_func2})}, we additionally introduce a forward loss as shown in Fig. \ref{fig_cycleGAN}(b) and a backward loss as shown in Fig. \ref{fig_cycleGAN}(c) to train the \textcolor{black}{CycleGAN}, each of which contains a consistency loss and a cycle-consistency loss.

We first evaluate the forward LS-GAN, which models the transmission process. The consistency loss is aimed at that the output estimated received signals should be consistent with the corresponding real ones, while the cycle-consistency loss is aimed at that, if using the output estimated received signals as the input of the backward LS-GAN, the new output should be consistent with the original real transmitted signals. Accordingly, we can formulate the forward loss as 
\begin{equation}
\label{GAN_func8}
{\mathcal{L}_{\rm fw}} = \alpha \Vert {\rm G_{s2y}}(\bm{s}) - \bm{y} \Vert_{1} + \beta \Vert {\rm G_{y2s}}({\rm G_{s2y}}(\bm{s})) - \bm{s} \Vert_{1} \text{,}
\end{equation}
where $\Vert\cdot\Vert_{1}$ is the $l1$-norm of the vector, \textcolor{black}{and $\alpha$, $\beta$ are the hyperparameters indicating the weights} of the consistency loss and the cycle-consistency loss, respectively. 

Analogously for the backward LS-GAN, which models the detector, \textcolor{black}{the consistency loss and the cycle-consistency loss are set under similar principles with converse input and output.} Thus the backward loss is given by
\begin{equation}
\label{GAN_func9}
{\mathcal{L}_{\rm bw}} = \gamma \Vert {\rm G_{y2s}}(\bm{y}) - \bm{s} \Vert_{1}  + \delta \Vert {\rm G_{s2y}}({\rm G_{y2s}}(\bm{y})) - \bm{y} \Vert_{1} \text{,}
\end{equation}
where $\gamma$ and $\delta$ are hyperparameters indicating the weights of the subordinated loss terms.

\subsection{Training Strategy\label{train_scheme}}
As illustrated in Fig. \ref{fig_train}, for each block of $K = P + D$ symbols, we conduct a supervised training phase in the \textcolor{black}{pilot-training period using pilots $[\bm{s}_{\rm P}, \bm{y}_{\rm P}]$ and then a semi-supervised training phase in the data-transmission period using both pilots $[\bm{s}_{\rm P}, \bm{y}_{\rm P}]$ and received payload data $\bm{y}_{\rm D}$.} In order to further augment the \textcolor{black}{training dataset}, let $[\bm{s}_{\rm P}^{(l)}, \bm{y}_{\rm P}^{(l)}]$ be the pilots in the $l$th block which is the current block for processing. Then the previous pilots $[\bm{s}_{\rm P}^{(l-1)}, \bm{y}_{\rm P}^{(l-1)}]$ in the $(l-1)$th block \textcolor{black}{are also used for the training} if the channel remains mostly unchanged. In fact since the channel is unknown at the receiver, whether the previous pilots are used actually depends on if they could make the model a progress \textcolor{black}{after augmented to the training dataset}. The model is updated block-by-block to keep tracking continuous block fading channels.    
\subsubsection{\textcolor{black}{Pilot-training} Period}
In the \textcolor{black}{pilot-training} period, \textcolor{black}{according to (\ref{GAN_func}) and (\ref{GAN_func2}), the optimal ${\rm D}_{\rm s2y}^{\ast}$, ${\rm D}_{\rm y2s}^{\ast}$, ${\rm G}_{\rm s2y}^{\ast}$, and ${\rm G}_{\rm y2s}^{\ast}$} are obtained by calculating the optimal network parameters ${\bm{\phi}^{\ast}_{\rm s2y}}$, ${\bm{\phi}^{\ast}_{\rm y2s}}$, ${\bm{\theta}^{\ast}_{\rm s2y}}$, and ${\bm{\theta}^{\ast}_{\rm y2s}}$ as follows:
\begin{equation}
\begin{split}
\label{train_obj1a}
{\bm{\phi}^{\ast}_{\rm s2y}} = &\arg\mathop{\min}\limits_{{\bm{\phi}_1}}\{\mathbb{E}_{\bm{s}, \bm{y}}[({\rm D}_{\rm s2y}([\bm{s}, \bm{y}]; \bm{\phi}_1) - 1)^{2} \\
& + ({\rm D}_{\rm s2y}([\bm{s}, {\rm G_{s2y}}(\bm{s})]; \bm{\phi}_1) + 1)^{2}]\} \text{,} 
\end{split}
\end{equation}
\begin{equation}
\begin{split}
\label{train_obj1b}
{\bm{\phi}^{\ast}_{\rm y2s}} = &\arg\mathop{\min}\limits_{{\bm{\phi}_2}}\{\mathbb{E}_{\bm{s}, \bm{y}}[({\rm D}_{\rm y2s}([\bm{y}, \bm{s}]; \bm{\phi}_2) - 1)^{2}\\
&+ ({\rm D}_{\rm y2s}([\bm{y}, {\rm G_{y2s}}(\bm{y})]; \bm{\phi}_2) + 1)^{2}]\} \text{,}
\end{split}
\end{equation}
\begin{equation}
\begin{split}
\label{train_obj1c}
[{\bm{\theta}^{\ast}_{\rm s2y}}, {\bm{\theta}^{\ast}_{\rm y2s}}] =& \arg\mathop{\min}\limits_{[{\bm{\theta}_{1}}, {\bm{\theta}_{2}}]}\{\mathbb{E}_{\bm{s}, \bm{y}}[{\rm D_{s2y}}([\bm{s}, {\rm G_{s2y}}(\bm{s}; \bm{\theta}_1)])^{2}\\
&+ {\rm D_{y2s}}([\bm{y}, {\rm G_{y2s}}(\bm{y}; \bm{\theta}_2)])^{2}\\
&+ \alpha_{1}\Vert {\rm G_{s2y}}(\bm{s}; \bm{\theta}_1) - \bm{y} \Vert_{1}\\
&+ \beta_{1}\Vert {\rm G_{y2s}}(\bm{y}; \bm{\theta}_2) - \bm{s} \Vert_{1}\\
&+ \gamma_{1}\Vert{\rm G_{y2s}}({\rm G_{s2y}}(\bm{s}; \bm{\theta}_1); \bm{\theta}_2) - \bm{s} \Vert_{1}\\
&+ \delta_{1}\Vert {\rm G_{s2y}}({\rm G_{y2s}}(\bm{y}; \bm{\theta}_2); \bm{\theta}_1) - \bm{y} \Vert_{1}]\} \text{,}
\end{split}
\end{equation}
where $\alpha_{1}$, $\beta_{1}$, $\gamma_{1}$, and $\delta_{1}$ are the hyperparameters indicating the weights of the subordinated loss terms defined in \mbox{Section \ref{cycleGAN}}. \textcolor{black}{During each training epoch, we first train ${\rm D_{s2y}}$ and ${\rm D_{y2s}}$ with fixed ${\rm G_{s2y}}$ and ${\rm G_{y2s}}$, and then train ${\rm G_{s2y}}$ and ${\rm G_{y2s}}$ with fixed ${\rm D_{s2y}}$ and ${\rm D_{y2s}}$.} Since the training purposes of the discriminators and generators are diametrically opposed to each other, these modules can keep guiding their opponents in an adversarial way, i.e., stronger generators can force the discriminator to capture deeper effects included in the transmission for stricter judgment, such as the precoding and nonlinear distortion, while stronger discriminators can force the generators to better model those effects, and then generate more accurate estimates. 

\begin{algorithm*}[!t]
\caption{Supervised Training in the \textcolor{black}{Pilot-training Period}}\label{supervised_training}
\begin{algorithmic}
\STATE 
\STATE $ \text{Divide the pilots } [\mathbf{S}_{\rm  P}^{(l)}, \mathbf{Y}_{\rm P}^{(l)}] \text{ into a training set } [\mathbf{S}_{\rm P}^{\rm train}, \mathbf{Y}_{\rm P}^{\rm train}] \text{ and a validation set } [\mathbf{S}_{\rm P}^{\rm val}, \mathbf{Y}_{\rm P}^{\rm val}] \text{.}$
\STATE $\text{Operate data augmentation.}$
\STATE $[\rm D_{s2y}^1, \rm D_{y2s}^1, \rm G_{s2y}^1, \rm G_{y2s}^1] \gets [\rm D_{s2y}, \rm D_{y2s}, \rm G_{s2y}, \rm G_{y2s}]\text{.}$
\STATE \textbf{Step 1:} $\text{Train } [\rm D_{s2y}, \rm D_{y2s}, \rm G_{s2y}, \rm G_{y2s}] \text{ by } [\mathbf{S}_{\rm P}^{\rm train}, \mathbf{Y}_{\rm P}^{\rm train}] \text{:}$
\STATE \hspace{0.5cm}\textbf {for} $ {\text{number of training iterations}} $ \textbf {do}
\STATE \hspace{1cm}$ \text{Sample } m \text{ \textcolor{black}{transmitted signals} } \{\bm{s}_{1}, \bm{s}_{2}, ..., \bm{s}_{m}\} \text{ in } \mathbf{S}_{\rm P}^{\rm train} \text{ and the \textcolor{black}{received signals} } \{\bm{y}_{1}, \bm{y}_{2}, ..., \bm{y}_{m}\} \text{ in } \mathbf{Y}_{\rm P}^{\rm train} \text{.}$
\STATE \hspace{1cm}$ \text{Update } {\rm D_{s2y}} \text{ by ascending the stochastic gradient:} $
\STATE $ $
\STATE \centerline{$ \nabla \frac{1}{m} \sum_{i=1}^m [({\rm D_{s2y}}([\bm{s}_{i}, \bm{y}_{i}]) - 1)^{2} + ({\rm D_{s2y}}([\bm{s}_{i}, {\rm G_{s2y}}(\bm{s}_{i})]) + 1)^{2}] \text{.}$}
\STATE \hspace{1cm}$ \text{Update } {\rm D_{y2s}} \text{ by ascending the stochastic gradient:} $
\STATE $ $
\STATE \centerline{$ \nabla \frac{1}{m} \sum_{i=1}^m [({\rm D_{y2s}}([\bm{y}_{i}, \bm{s}_{i}]) - 1)^{2} + ({\rm D_{y2s}}([\bm{y}_{i}, {\rm G_{y2s}}(\bm{y}_{i})]) + 1)^{2}] \text{.}$}
\STATE \hspace{1cm}$ \text{Update } {\rm G_{s2y}} \text{ and } {\rm G_{y2s}} \text{ by ascending the stochastic gradient:} $
\STATE $ $
\STATE \centerline{$ \nabla \frac{1}{m} \sum_{i=1}^m [{\rm D_{s2y}}([\bm{s}_{i}, {\rm G_{s2y}}(\bm{s}_{i})])^{2} + {\rm D_{y2s}}([\bm{y}_{i}, {\rm G_{y2s}}(\bm{y}_{i})])^{2} + \alpha_{1}\Vert {\rm G_{s2y}}(\bm{s}_{i}) - \bm{y}_{i} \Vert_{1}$}
\STATE \centerline{$ + \beta_{1}\Vert {\rm G_{y2s}}(\bm{y}_{i}) - \bm{s}_{i} \Vert_{1} + \gamma_{1}\Vert{\rm G_{y2s}}({\rm G_{s2y}}(\bm{s}_{i})) - \bm{s}_{i} \Vert_{1} + \delta_{1}\Vert {\rm G_{s2y}}({\rm G_{y2s}}(\bm{y}_{i})) - \bm{y}_{i} \Vert_{1}] \text{.}$}
\STATE \hspace{1cm}$ \text{Check the BER performance of } {\rm G_{y2s}} \text{ in } [\mathbf{S}_{\rm P}^{\rm val}, \mathbf{Y}_{\rm P}^{\rm val}] \text{.}$ 
\STATE \hspace{0.5cm}\textbf {end for}
\STATE \textbf{Step 2:} $\text{Train } [\rm D_{s2y}^1, \rm D_{y2s}^1, \rm G_{s2y}^1, \rm G_{y2s}^1] \text{ by } [(\mathbf{S}_{\rm P}^{\rm train}, \mathbf{S}_{\rm  P}^{(l-1)}), (\mathbf{Y}_{\rm P}^{\rm train}, \mathbf{Y}_{\rm P}^{(l - 1)})] \text{ following the same strategy in}$ \textbf{Step 1}$.$
\STATE \textbf {if } ${\rm G_{y2s}^{1}} \text{ performs better than } {\rm G_{y2s}} \text{:}$
\STATE \hspace{0.5cm} $[\rm D_{s2y}, \rm D_{y2s}, \rm G_{s2y}, \rm G_{y2s}] \gets [\rm D_{s2y}^1, \rm D_{y2s}^1, \rm G_{s2y}^1, \rm G_{y2s}^1]\text{.}$
\STATE \hspace{0.5cm} $[\mathbf{S}_{\rm P}^{\rm train}, \mathbf{Y}_{\rm P}^{\rm train}] \gets [(\mathbf{S}_{\rm P}^{\rm train}, \mathbf{S}_{\rm  P}^{(l - 1)}), (\mathbf{Y}_{\rm P}^{\rm train}, \mathbf{Y}_{\rm P}^{(l - 1)})]\text{.}$
\STATE \textbf {end if}
\vspace{0.1cm}
\end{algorithmic}
\label{alg1}
\end{algorithm*}

\begin{algorithm*}[!t]
\caption{Semi-Supervised Training in the \textcolor{black}{Data-transmission Period}}\label{semi-supervised_training}
\begin{algorithmic}
\STATE 
\STATE $ \tilde{\mathbf{S}}_{\rm D} \gets {\rm G_{y2s}}(\mathbf{Y}_{\rm D}) \text{.}$
\STATE $[\mathbf{S}^{\rm train}, \mathbf{Y}^{\rm train}] \gets [(\mathbf{S}_{\rm P}^{\rm train}, \tilde{\mathbf{S}}_{\rm D}), (\mathbf{Y}_{\rm P}^{\rm train}, \mathbf{Y}_{\rm D})]\text{.}$
\STATE \textbf {for} $ {\text{number of training epochs}} $ \textbf {do}
\STATE \hspace{0.5cm}$ \text{Sample } m \text{ \textcolor{black}{transmitted signals} } \{\bm{s}_{1}, \bm{s}_{2}, ..., \bm{s}_{m}\} \text{ in } \mathbf{S}^{\rm train} \text{ and the \textcolor{black}{received signals} } \{\bm{y}_{1}, \bm{y}_{2}, ..., \bm{y}_{m}\} \text{ in } \mathbf{Y}^{\rm train} \text{.}$
\STATE \hspace{0.5cm}$ \text{Update } {\rm D_{s2y}} \text{ by ascending the stochastic gradient:} $
\STATE $ $
\STATE \centerline{$ \nabla \frac{1}{m} \sum_{i=1}^m [({\rm D_{s2y}}([\bm{s}_{i}, \bm{y}_{i}]) - 1)^{2} + ({\rm D_{s2y}}([\bm{s}_{i}, {\rm G_{s2y}}(\bm{s}_{i})]) + 1)^{2}] \text{.}$}
\STATE \hspace{0.5cm}$ \text{Update } {\rm D_{y2s}} \text{ by ascending the stochastic gradient:} $
\STATE $ $
\STATE \centerline{$ \nabla \frac{1}{m} \sum_{i=1}^m [({\rm D_{y2s}}([\bm{y}_{i}, \bm{s}_{i}]) - 1)^{2} + ({\rm D_{y2s}}([\bm{y}_{i}, {\rm G_{y2s}}(\bm{y}_{i})]) + 1)^{2}] \text{.}$}
\STATE \hspace{0.5cm}$ \text{Update } {\rm G_{s2y}} \text{ and } {\rm G_{y2s}} \text{ by ascending the stochastic gradient:} $
\STATE $ $
\STATE \centerline{$ \nabla \frac{1}{m} \sum_{i=1}^m [{\rm D_{s2y}}([\bm{s}_{i}, {\rm G_{s2y}}(\bm{s}_{i})])^{2} + {\rm D_{y2s}}([\bm{y}_{i}, {\rm G_{y2s}}(\bm{y}_{i})])^{2} + \alpha_{2}\Vert {\rm G_{s2y}}(\bm{s}_{i}) - \bm{y}_{i} \Vert_{1}$}
\STATE \centerline{$ + \beta_{2}\Vert {\rm G_{y2s}}(\bm{y}_{i}) - \bm{s}_{i} \Vert_{1} + \gamma_{2}\Vert{\rm G_{y2s}}({\rm G_{s2y}}(\bm{s}_{i})) - \bm{s}_{i} \Vert_{1} + \delta_{2}\Vert {\rm G_{s2y}}({\rm G_{y2s}}(\bm{y}_{i})) - \bm{y}_{i} \Vert_{1}] \text{.}$}
\STATE \hspace{0.5cm}$ \text{Check the BER performance of } {\rm G_{y2s}} \text{ in } [\mathbf{S}_{\rm P}^{\rm val}, \mathbf{Y}_{\rm P}^{\rm val}] \text{.}$ 
\STATE \hspace{0.5cm}$ \text{If progress of } {\rm G_{y2s}} \text{ is made:}$
\STATE \hspace{1cm}$ \tilde{\mathbf{S}}_{\rm D} \gets {\rm G_{y2s}}(\mathbf{Y}_{\rm D}) \text{.}$
\STATE \hspace{1cm}$[\mathbf{S}^{\rm train}, \mathbf{Y}^{\rm train}] \gets [(\mathbf{S}_{\rm P}^{\rm train}, \tilde{\mathbf{S}}_{\rm D}), (\mathbf{Y}_{\rm P}^{\rm train}, \mathbf{Y}_{\rm D})]\text{.}$
\STATE \textbf {end for}
\vspace{0.1cm}
\end{algorithmic}
\label{alg2}
\end{algorithm*}

In fact, before the training we also conduct some pre-processing of the dataset, such as data augmentation and normalization, to improve the stability and efficiency of learning. Implementation details and empirical training \textcolor{black}{setups} will be further elaborated in Section \ref{implementation}. Moreover, to make use of the previous pilots in the last block, we actually train the model twice parallelly. One uses the current pilots only and the other uses pilots from both the current block and the last block. After training we compare the performance of the two temporary models and save the better one. Then the saved model is set as the initial value for subsequent semi-supervised training in the \textcolor{black}{data-transmission} period. The previous pilots are kept for training if they make the model a progress and are discarded \textcolor{black}{otherwise}. Accordingly, for the $l$th block, the detailed supervised training algorithm \textcolor{black}{in the pilot-training period} is shown in \mbox{\textbf{Algorithm \ref{alg1}}}, where $m$ is the batch size.

\subsubsection{\textcolor{black}{Data-transmission} Period}
In the \textcolor{black}{data-transmission} period, we propose a semi-supervised training strategy to update the model. Before the training, we firstly use the initial ${\rm G_{y2s}}$ to obtain a preliminary estimation of $\bm{s}_{\rm D}$\textcolor{black}{, denoted by $\tilde{\bm{s}}_{\rm D}$,} for each single received payload $\bm{y}_{\rm D}$ as
\begin{equation}
\label{train_temp}
\tilde{\bm{s}}_{\rm D} = {\rm G_{y2s}}(\bm{y}_{\rm D}) \text{.}
\end{equation}
Then the data pair of $[\tilde{\bm{s}}_{\rm D}, \bm{y}_{\rm D}]$ is used to augment the training set and \textcolor{black}{the updated ${\rm D}_{\rm s2y}^{\ast}$, ${\rm D}_{\rm y2s}^{\ast}$, ${\rm G}_{\rm s2y}^{\ast}$, and ${\rm G}_{\rm y2s}^{\ast}$} are obtained by updating the optimal network parameters ${\bm{\phi}^{\ast}_{\rm s2y}}$, ${\bm{\phi}^{\ast}_{\rm y2s}}$, ${\bm{\theta}^{\ast}_{\rm s2y}}$ and ${\bm{\theta}^{\ast}_{\rm y2s}}$ as follows:
\begin{equation}
\begin{split}
\label{train_obj2a}
{\bm{\phi}^{\ast}_{\rm s2y}} = &\arg\mathop{\min}\limits_{\bm{\phi}_{1}}\{\mathbb{E}_{\bm{y}}[({\rm D_{s2y}}([\tilde{\bm{s}}, \bm{y}]; \bm{\phi}_{1}) - 1)^{2} \\
& + ({\rm D_{s2y}}([\tilde{\bm{s}}, {\rm G_{s2y}}(\tilde{\bm{s}})]; \bm{\phi}_{1}) + 1)^{2}]\} \text{,} 
\end{split}
\end{equation}
\begin{equation}
\begin{split}
\label{train_obj2b}
{\bm{\phi}^{\ast}_{\rm y2s}} = &\arg\mathop{\min}\limits_{\bm{\phi}_{2}}\{\mathbb{E}_{\bm{y}}[({\rm D_{y2s}}([\bm{y}, \tilde{\bm{s}}]; \bm{\phi}_{2}) - 1)^{2}\\
&+ ({\rm D_{y2s}}([\bm{y}, {\rm G_{y2s}}(\bm{y})]; \bm{\phi}_{2}) + 1)^{2}]\} \text{,}
\end{split}
\end{equation}
\begin{equation}
\begin{split}
\label{train_obj2c}
[{\bm{\theta}^{\ast}_{\rm s2y}}, {\bm{\theta}^{\ast}_{\rm y2s}}] =& \arg\mathop{\min}\limits_{[{\bm{\theta}_{1}}, {\bm{\theta}_{2}}]}\{\mathbb{E}_{\bm{y}}[{\rm D_{s2y}}([\tilde{\bm{s}}, {\rm G_{s2y}}(\tilde{\bm{s}}; \bm{\theta}_{1})])^{2}\\
&+ {\rm D_{y2s}}([\bm{y}, {\rm G_{y2s}}(\bm{y}; \bm{\theta}_{2})])^{2}\\
&+ \alpha_{2}\Vert {\rm G_{s2y}}(\tilde{\bm{s}}; \bm{\theta}_{1}) - \bm{y} \Vert_{1}\\
&+ \beta_{2}\Vert {\rm G_{y2s}}(\bm{y}; \bm{\theta}_{2}) - \tilde{\bm{s}} \Vert_{1}\\
&+ \gamma_{2}\Vert{\rm G_{y2s}}({\rm G_{s2y}}(\tilde{\bm{s}}; \bm{\theta}_{1}); \bm{\theta}_{2}) - \tilde{\bm{s}} \Vert_{1}\\
&+ \delta_{2}\Vert {\rm G_{s2y}}({\rm G_{y2s}}(\bm{y}; \bm{\theta}_{2}); \bm{\theta}_{1}) - \bm{y} \Vert_{1}]\} \text{,}
\end{split}
\end{equation}
where $\alpha_{2}$, $\beta_{2}$, $\gamma_{2}$, and $\delta_{2}$ are hyperparameters indicating the weights of the subordinated loss terms. Similar to the supervised learning phase, \textcolor{black}{during each training epoch, we first train ${\rm D_{s2y}}$ and ${\rm D_{y2s}}$ with fixed ${\rm G_{s2y}}$ and ${\rm G_{y2s}}$, and then train ${\rm G_{s2y}}$ and ${\rm G_{y2s}}$ with fixed ${\rm D_{s2y}}$ and ${\rm D_{y2s}}$.} Note that $\tilde{\bm{s}}_{\rm D}$ is not real data, which contains some error bits that may mislead the model \textcolor{black}{and thus bring instability to the training}. Nevertheless, such training strategy still makes sense since the error bits normally share a relatively small percentage \textcolor{black}{due to previous training in the pilot-training period. In order to further avoid the misguidance of error bits, we keep updating $\tilde{\bm{s}}_{\rm D}$ whenever the model makes a progress under the validation dataset so that the number of error bits keeps decreasing during the training.} Accordingly, the detailed semi-supervised training algorithm \textcolor{black}{in the data-transmission period} is shown in \textbf{Algorithm \ref{alg2}}.

\begin{table*}[!t]
\renewcommand\arraystretch{1.5}
\caption{\textcolor{black}{An Overview of Network Configurations and Parameters}\label{layout}}
\centering
\begin{tabular}{| c | c | c | c | c |}
\hline
& ${\rm G_{s2y}}$ & ${\rm G_{y2s}}$ & ${\rm D_{s2y}}$ & ${\rm D_{y2s}}$\\
\hline
Input layer & 16 & 16 & 32 & 32\\
\hline
Layer 1 & Dense 256 & Dense 256 & Dense 512 & Dense 512\\
\hline
Layer 2 & LeakyRelu($\alpha = 0.2$) & LeakyRelu($\alpha = 0.2$) & LeakyRelu($\alpha = 0.2$) & LeakyRelu($\alpha = 0.2$)\\
\hline 
Layer 3 & Dropout(0.1) & Dropout(0.1) & Dropout(0.1) & Dropout(0.1)\\ 
\hline
Layer 4 & Dense 512 & Dense 512 & Dense 256 & Dense 256\\
\hline
Layer 5 & LeakyRelu($\alpha = 0.2$) & LeakyRelu($\alpha = 0.2$) & LeakyRelu($\alpha = 0.2$) & LeakyRelu($\alpha = 0.2$)\\
\hline 
Layer 6 & Dropout(0.1) & Dropout(0.1) & Dropout(0.1) & Dropout(0.1)\\
\hline
Output layer & Dense 16-tanh & Dense 16-tanh & Dense 1 & Dense 1\\
\hline
\end{tabular}
\end{table*}

The proposed training strategy leads the model to learn a bidirectional mapping relationship between the transmitted signal and the received signal. It makes sufficient use of all the current pilots, previous pilots and the received payload data. Since no prior assumption of the mapping relationship is made, the generators are guided only by the discriminators and the consistency of data, thus promising better performance if the underlying channel model is unknown. After the whole training process is completed, the ${\rm G_{s2y}}$ is considered as a DNN model of the forward transmission process and the ${\rm G_{y2s}}$ is used as the signal detector to \textcolor{black}{recover the desired payload data.}  

\subsection{Computational Complexity Analysis \label{complexity}}
Let $n$ be the length of the payload data. Then the computational complexity of \textcolor{black}{a} classic LMMSE detector is $O(n)$. For the \textcolor{black}{CycleGAN} detector, the forward-pass computational complexity of the detector ${\rm G_{y2s}}$ is also $O(n)$. Note that beside the forward-pass calculation, the actual time cost also depends on the training process and the time complexity of the \textcolor{black}{CycleGAN} model is $O(N_{\rm r}^2 n^2)$. To overcome this drawback, we set the model trained in the last block as the initial value for the next block, so that the previous knowledge \textcolor{black}{is kept for the subsequent} learning. Generally in our experiments for each block the required number of training epochs is no more than 5000 and for successive blocks that the channel remains unchanged, the model trains about twice faster.

\section{\textcolor{black}{Implementation Details} \label{implementation}}
In this \textcolor{black}{section,} we introduce the implementation details of the \textcolor{black}{CycleGAN} detector. Our experiments are conducted under conda 4.9.2, python 3.8.5, and pytorch 1.11.0, where python is the coding language, conda is the environment for implementations of functions, pytorch is a common framework to develop DL models. In our experiments we set $N_{\rm s} = 64$, $N_{\rm r} = 8$, $K = 320$, $P = 64$, and $D = 256$.

\subsection{Network Parameters}

\subsubsection{Generators}
${\rm G_{s2y}}$ and ${\rm G_{y2s}}$ are DNNs including \textcolor{black}{two} dense hidden layers and \textcolor{black}{one} dense output layer, respectively. The input layer has 16 neurons. The first hidden layer has 256 neurons and the second hidden layer has 512 neurons. Each hidden layer follows an activation layer of \textit{`LeakyRelu'}, with $\alpha = 0.2$ as the slope in negative-half-axis. Besides, we set a dropout layer at the end of each hidden layer to resist \mbox{overfitting \cite{ref25, ref26}}, with the dormancy rate of 0.1. The output layer has 16 neurons, after which follows an activation layer of \textit{`tanh'}. All the network parameters are regularized by \textcolor{black}{an $l2$} regularizer.

\subsubsection{Discriminators}
${\rm D_{s2y}}$ and ${\rm D_{y2s}}$ are DNNs including \textcolor{black}{two} dense hidden layers and \textcolor{black}{one} dense output layer, respectively. The input layer has 32 neurons. The first hidden layer has 512 neurons and the second hidden layer has 256 neurons. Each hidden layer follows an activation layer of \textit{`LeakyRelu'}, with $\alpha = 0.2$ as the slope in negative-half-axis. Besides, we set a dropout layer at the end of each hidden layer to resist overfitting, with the dormancy rate of 0.1. The output layer has \textcolor{black}{one} neuron, the discriminator of LS-GAN \textcolor{black}{has no} activation layer at the output side\cite{ref22}. All the network parameters are regularized by \textcolor{black}{an $l2$} regularizer. 

\textcolor{black}{The layout of the four networks are given in \mbox{Table \ref{layout}}}. After setting up the four modules, we \textcolor{black}{incorporate} them for the \textcolor{black}{CycleGAN} proposed in Section \ref{cycleGAN}. 

\subsection{\textcolor{black}{Pre-processing}}
Pre-processing of the dataset is necessary to ensure a stable training. Our pre-processing mainly includes data augmentation, signal flattening, and normalization.
\subsubsection{Data Augmentation}
Note that the scale of pilots is limited, which may not be enough for training the model and indicates a high risk of overfitting. Thus, we consider about adding white noise for data augmentation to obtain a larger training set and a larger validation set.

The mapping of data augmentation by adding white noise is given by
\begin{equation}
\begin{cases}
\label{preprocessA}
\bm{s}_{\rm aug} = \bm{s} + \bm{n}_{s} \\
\bm{y}_{\rm aug} = \bm{y} + \bm{n}_{y} \text{,}
\end{cases}
\end{equation}
where $\bm{s}_{\rm aug}$ and $\bm{y}_{\rm aug}$ are the augmented signals, $\bm{s}$ and $\bm{y}$ are the real signals, and $\bm{n}_{s}$ and $\bm{n}_{y}$ are the noise added for augmentation.

\textcolor{black}{For each block we divide the pilot sequence into a training dataset a validation dataset by 3 to 1, and then augment each dataset by 10 times for supervised learning. The payload data is augmented by 5 times and is packed with the pilot training dataset for semi-supervised learning. The original payload data is used as the test dataset.}

\subsubsection{Signal Flattening}
\textcolor{black}{Since it is computationally complex for neural networks to operate plural calculation in training, we flattened the complex signals into real signals by separating the real part and the imaginary part, the mapping of flattening is given by}
\begin{equation}
\begin{cases}
\label{preprocessB}
\bm{s}_{i,2k}^{\rm flt} = {\rm Re}(\bm{s}_{i,k})\\
\bm{s}_{i, 2k+1}^{\rm flt} = {\rm Im}(\bm{s}_{i,k})\\
\bm{y}_{i, 2k}^{\rm flt} = {\rm Re}(\bm{y}_{i,k})\\
\bm{y}_{i, 2k+1}^{\rm flt} = {\rm Im}(\bm{y}_{i,k}) \text{,}
\end{cases}
\end{equation}
where $\bm{s}_{i,2k}^{\rm flt}$ and $\bm{s}_{i, 2k+1}^{\rm flt}$ are the flattened transmitted signals, $\bm{y}_{i, 2k}^{\rm flt}$ and $\bm{y}_{i, 2k+1}^{\rm flt}$ are the flattened received signals, ${\rm Re}(\cdot)$, ${\rm Im}(\cdot)$ are the real part and the imaginary part of the plural respectively, $\bm{s}_{i,k}$, $\bm{y}_{i,k}$ are the original signals, $i$ is the index of frame, and $k$ is the index of position. After the signal flattening, the mapping between complex signals is transformed into the mapping between real signals, which is feasible to be learned by neural networks. 

\subsubsection{Normalization}
Typically, while constructing a neural network, an activation function is set at the output side of each layer to introduce nonlinearity so that the network could approach complex nonlinear mapping relationships. \textcolor{black}{Since general activation functions are sensitive only for input that is close to zero, we normalized the flattened signals to set the numeric values between $[-1, 1]$. The mapping of normalization is given by}
\begin{equation}
\begin{cases}
\label{preprocessC}
\bm{s}_{i, k}^{\rm norm} = \frac{\bm{s}_{i, k}^{\rm flt}}{\mathop{\max}\limits_{1\leq n \leq K}\{\vert \bm{s}_{n, k}^{\rm flt} \vert\}}\\
\bm{y}_{i, k}^{\rm norm} = \frac{\bm{y}_{i, k}^{\rm flt}}{\mathop{\max}\limits_{1\leq n \leq K}\{\vert \bm{s}_{n, k}^{\rm flt} \vert\}} \text{,}
\end{cases}
\end{equation}
where $\bm{s}_{i, k}^{\rm norm}$ is the normalized transmitted signal, $\bm{y}_{i, k}^{\rm norm}$ is the normalized received signal, $i$ is the index of frame, and $k$ is the index of position. 

By such pre-processing \textcolor{black}{procedures,} we modify the dataset into a more regular form to ensure a stable and efficient training procedure.    
\subsection{Training Procedure}
After pre-processing, we then start training using the processed dataset, following the training strategy proposed in \mbox{Section \ref{train_scheme}}. To improve the efficiency of training, we introduce some empirical training \textcolor{black}{setups} for implementation, including batch training, label invert, and early stopping. 

\subsubsection{Batch Training}
To conduct the gradient-based updates, in our experiments we use Adam as the optimizer with the learning rate of 0.0002 and exponential decay rate of (0.5, 0.99). Note that the variance of data can be quite large if we feed the whole dataset to train the model, which will cause inaccurate gradient calculation. Therefore, in order to obtain more accurate gradient, we set batch size $m = 128$ and feed a batch of data to the model rather than the whole dataset for each training iteration.

\subsubsection{Label Invert}
Sometimes the model may get stuck in local-optimal solutions if it erroneously learns some superficial features. Therefore, in practice we set a probability of 5\% in each training iteration to invert the data labels, \textcolor{black}{i.e.,} \textcolor{black}{to set real signals as fake and set generated signals as real. In this way we help the model to recheck the acquired knowledge and modify the learning direction.}

\subsubsection{\textcolor{black}{Early Stopping}}
Unlike \textcolor{black}{conventional iterative algorithms which have categorical stopping conditions, the training of \textcolor{black}{CycleGAN} can be terminated depending on the performance of ${\rm G_{y2s}}$. If for several consecutive training epochs the detection performance of ${\rm G_{y2s}}$ remains unimproved, we stop the training to save time and avoid further overfitting.} In our experiments we set the early stopping patience as 100 epochs.  

The parameter setup in the model training phase is summarized in Table \ref{training para}.     

\begin{table*}[!t]
\renewcommand\arraystretch{1.5}
\caption{\textcolor{black}{Training Parameters}\label{training para}}
\centering
\begin{tabular}{| c | c | c | c | c | c |}
\hline
Training optimizer & Learning rate & Exponential decay rate & Batch size & Probability of label invert & Early stopping patience\\
\hline
Adam & 0.0002 & (0.5, 0.99) & 128 & 5\% & 100 epochs\\
\hline
\end{tabular}
\end{table*}

\begin{figure*}[t]
\centering
\subfloat[BER performance without distortion.]{
    \centering
    \begin{minipage}[t]{3.4in}
        \centering       
        \includegraphics[width=\linewidth]{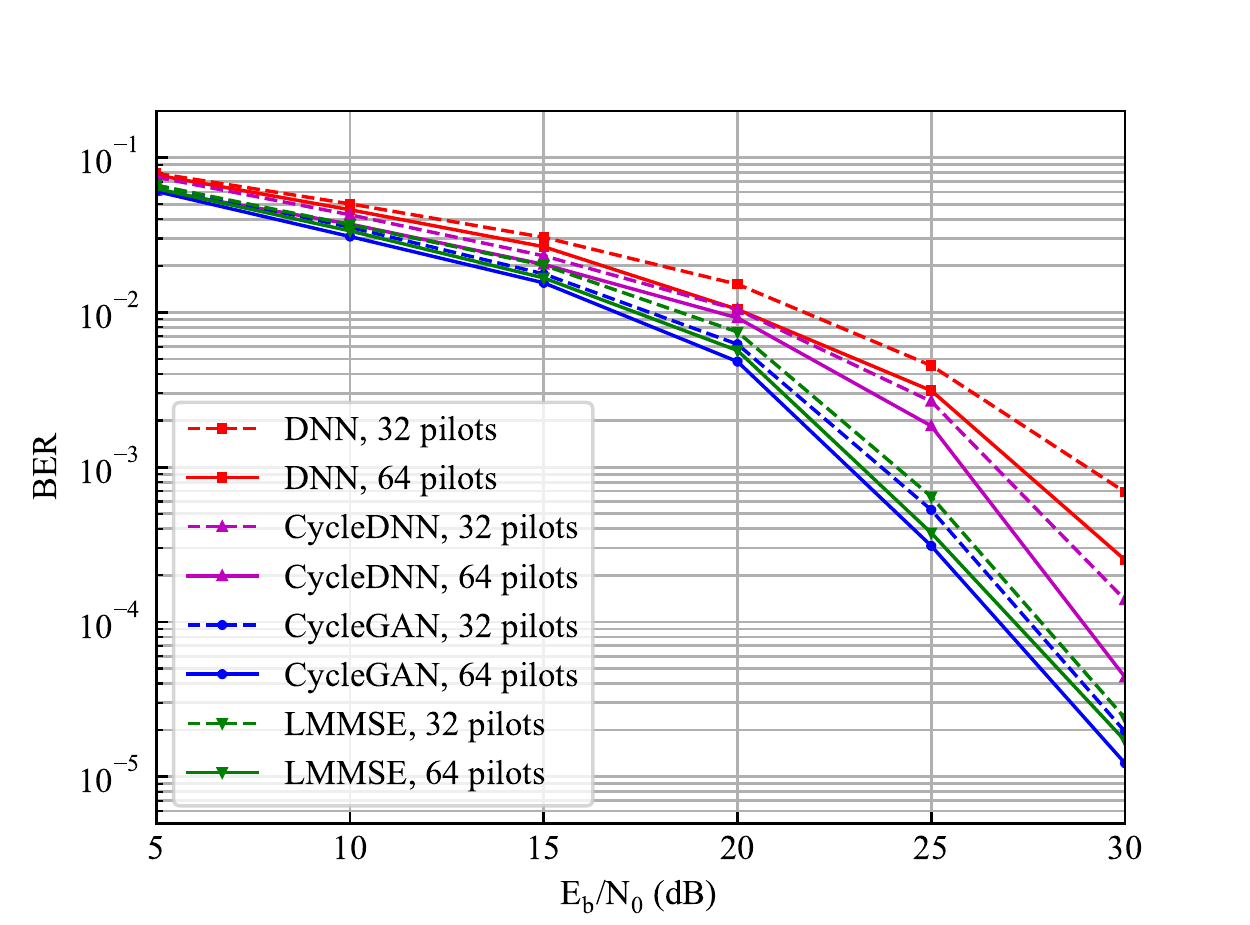}
        \end{minipage}
}%
\subfloat[Achievable rate without distortion.]{
    \centering
    \begin{minipage}[t]{3.4in}
        \centering       
        \includegraphics[width=\linewidth]{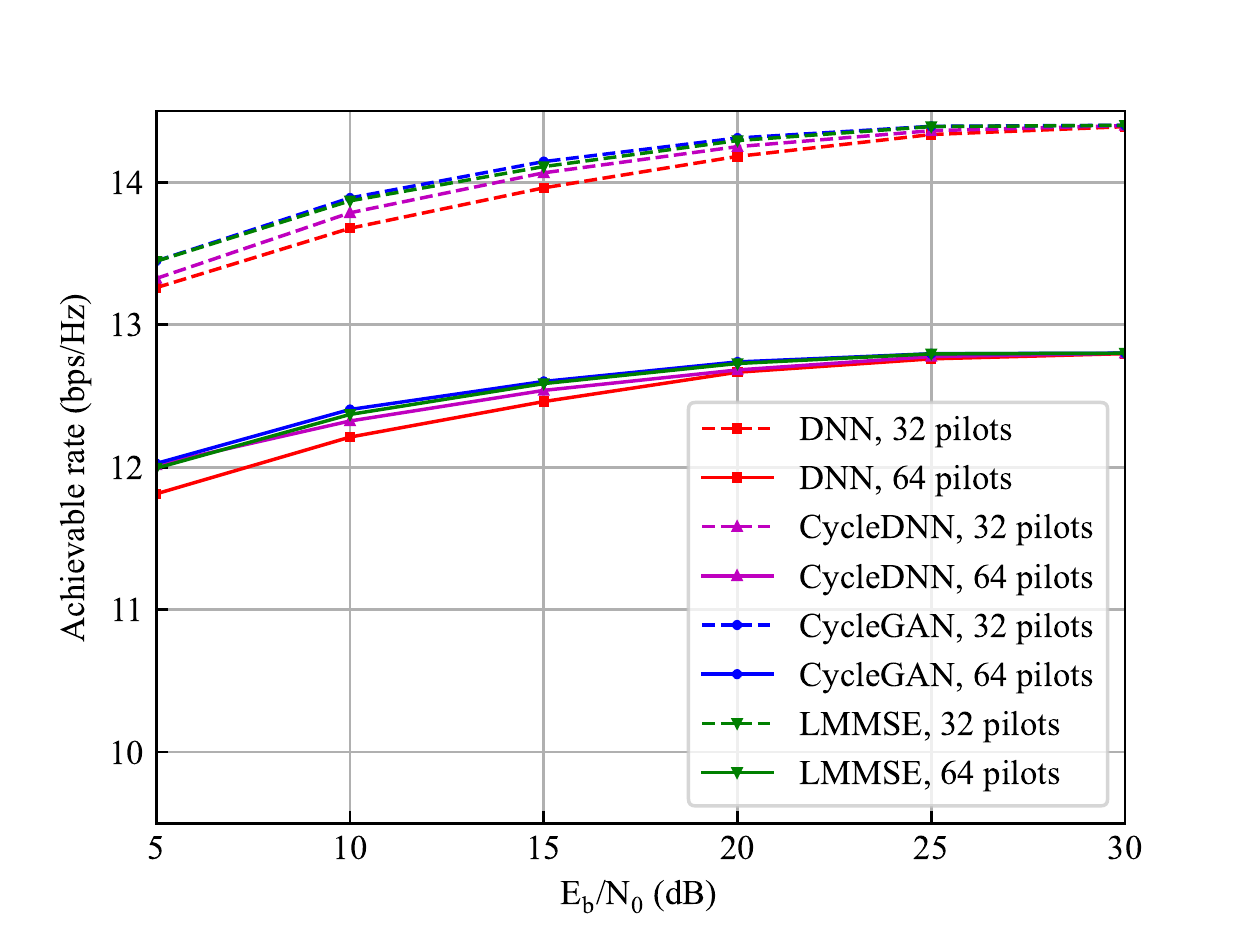}
    \end{minipage}
}%
\caption{\textcolor{black}{(a) BER and (b) achievable rate performance of the semi-blind \textcolor{black}{CycleGAN} detector, the semi-blind DNN detector, the semi-blind \textcolor{black}{CycleDNN} detector and the non-blind LMMSE without nonlinear distortion.}}
\label{fig_no_nonlinear}
\end{figure*}

\section{\textcolor{black}{Numerical Results}}
\subsection{Simulation Setup \label{setup}}
\textcolor{black}{To evaluate the performance of the proposed \textcolor{black}{CycleGAN} detector, we first consider the downlink transmission under Rayleigh block fading channel.} The channel is generated by the \textcolor{black}{Jakes'} Model \cite{ref27}. \textcolor{black}{Detailed channel generation process is given in the Appendix \ref{channel generation}. Then we} assume that the CSI is perfectly estimated at the transmitter so that a precoder based on singular value decomposition (SVD) can be implemented. To assess the robustness of the detectors against nonlinear distortion, nonlinearity of \textcolor{black}{PA} at the transmitter is also introduced. We model the nonlinearity by $g(\bm{s})$ as proposed in \cite{ref26}:
\begin{equation}
\label{nonlinearity}
g(\bm{s}) = \sum_{i=1}^{N}{a_{2i-1} \bm{s}^{2i-1}} \text{.}
\end{equation}   
Quadrature phase shift keying (QPSK) is adopted for modulation and the addictive noise is assumed white Gaussian noise. 

Besides the \textcolor{black}{CycleGAN} detector, we also setup a typical LMMSE detector \cite{ref10}, a basic DNN detector \cite{ref29}, and a \textcolor{black}{CycleDNN} detector for comparison. The LMMSE detector is supposed to be non-blind, \textcolor{black}{for which the SVD of the channel is also perfectly known at the receiver so that the equivalent diagonal channel matrix could be estimated for detection.} Other detectors are supposed to be semi-blind, for which there is no extra prior knowledge except \textcolor{black}{for} the pilots, and no pre-decoder at the receiver side. The \textcolor{black}{CycleDNN} detector is set by coupling only ${\rm G_{y2s}}$ and ${\rm G_{s2y}}$ into a bidirectional loop, which follows a semi-supervised training strategy similar to the \textcolor{black}{CycleGAN} detector without optimizing the discriminators. The DNN detector has the same architecture as the ${\rm G_{y2s}}$ but \textcolor{black}{since there is no bidirectional loop, the cycle-consistency loss is excluded in the objective function.} \textcolor{black}{All the modules of the communication system in our simulations are implemented in Python.}   

\begin{figure*}[t]
\centering
\subfloat[BER performance with distortion.]{
    \centering
    \begin{minipage}[t]{3.4in}
        \centering       
        \includegraphics[width=\linewidth]{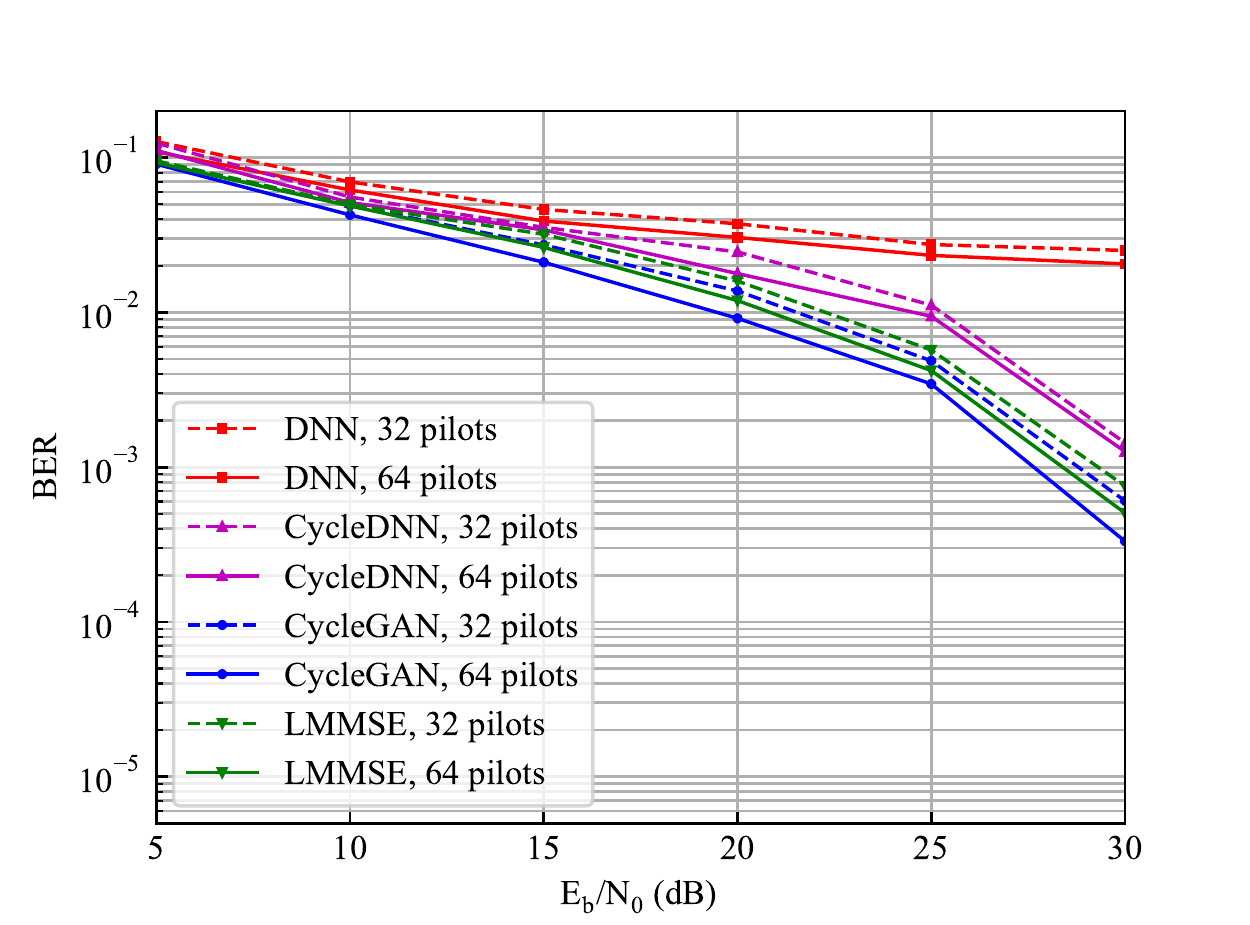}
        \end{minipage}
}%
\subfloat[Achievable rate with distortion.]{
    \centering
    \begin{minipage}[t]{3.4in}
        \centering       
        \includegraphics[width=\linewidth]{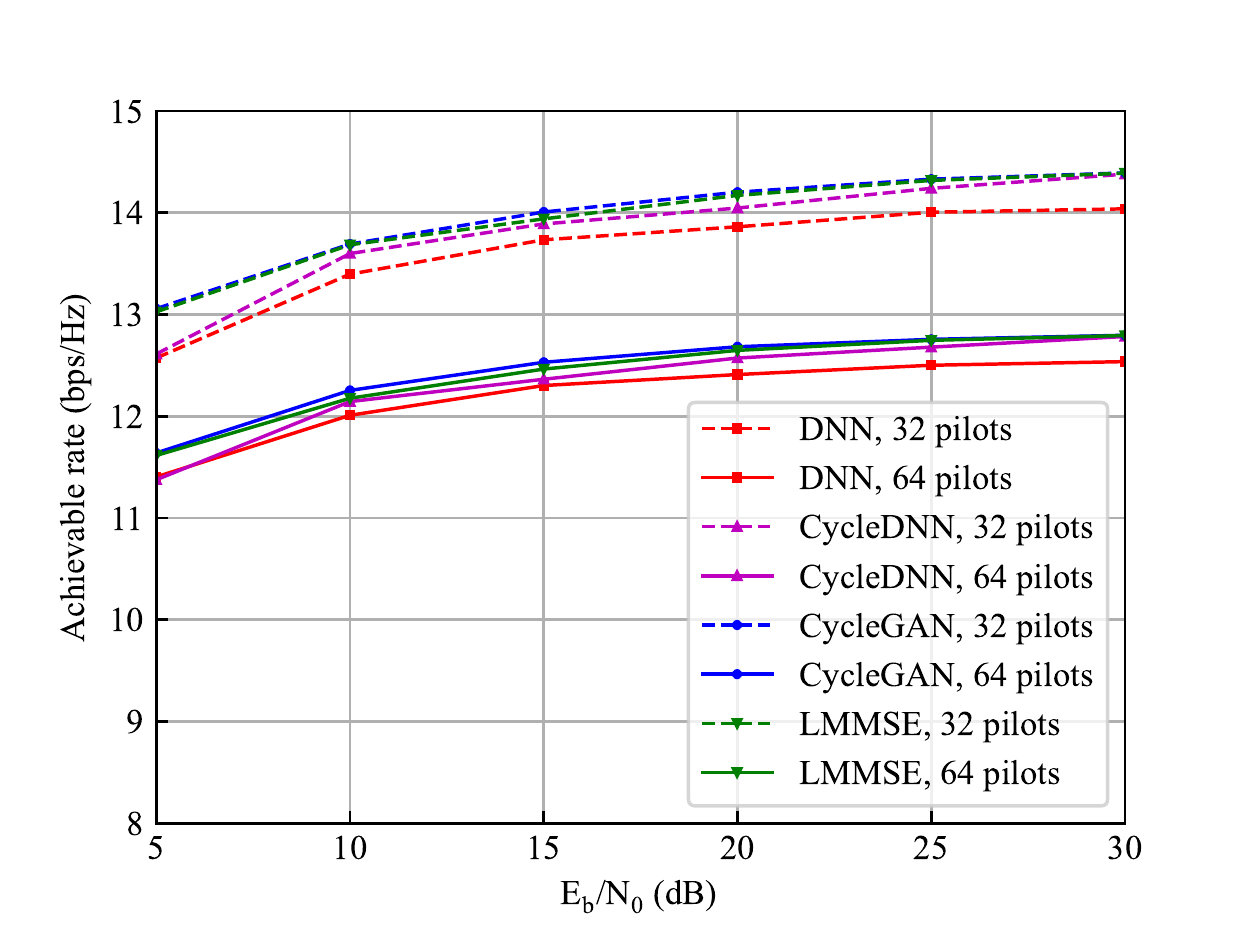}
    \end{minipage}
}%
\caption{\textcolor{black}{(a) BER and (b) achievable rate performance of the semi-blind \textcolor{black}{CycleGAN} detector, the semi-blind DNN detector, the semi-blind \textcolor{black}{CycleDNN} detector and the non-blind LMMSE with nonlinear distortion.}}
\label{fig_nonlinear}
\end{figure*}

\subsection{\textcolor{black}{Detection Performance} without Nonlinear Distortion \label{exp1}}
Firstly, \textcolor{black}{we evaluate the BER and achievable rate performance of the four detectors under transmissions without nonlinear distortion. We compare the performance of the \textcolor{black}{CycleGAN} detector against other benchmarks with bit signal-to-noise ratio $\rm (E_b/N_0)$ from 5 dB to 30 dB. For each $\rm E_b/N_0$ value, we generated 100 successive blocks of signals randomly, each block containing 64 pilot symbols and 256 symbols of payload data (\textcolor{black}{i.e.,} the overhead is 20\%). We also set up \textcolor{black}{another} simulation with 32 pilot symbols to investigate the instability introduced by pilot length. For each block we record the average BER and achievable rate of the detectors respectively for comparison as an assessment.} 

Experiments demonstrate that for $\rm E_b/N_0$ from 5 dB to 30 dB, \textcolor{black}{as illustrated in Fig. \ref{fig_no_nonlinear}, the BER and achievable rate performance of the \textcolor{black}{CycleGAN} detector stays better than other detectors and the gap between the \textcolor{black}{CycleGAN} detector and other detectors keeps expanding.} The DNN detector performs the worst since it falls into severe overfitting due to the limitation of pilots. The \textcolor{black}{CycleDNN} detector performs much better than the DNN detector but since there is no guidance of discriminators, it still can not accurately model the implicit relationship between signals, such as the precoding process, thus outperformed by the non-blind LMMSE detector. The performance of the non-blind LMMSE detector is close to that of the \textcolor{black}{CycleGAN} detector since there is no nonlinear distortion in the transmission and the non-blind LMMSE detector has perfect knowledge of CSI. \textcolor{black}{The decrease in pilot length leads to worse detection performance but the training remains stable.}  

\subsection{Detection Performance with Nonlinear Distortion \label{exp2}}
\textcolor{black}{Since} the LMMSE detector is based on linear algorithm which may face difficulty when processing transmissions under complex scenarios, we then evaluate the robustness of the detectors against nonlinear distortion. In our experiments we set $g(\bm{s})$ in \textcolor{black}{Section} \ref{setup} with $N = 3$, $a_1 = 1$, $a_3 = -1.5$, and $a_5 = -0.3$. Other simulation configurations are the same as in \textcolor{black}{Section} \ref{exp1}.  

\textcolor{black}{The BER and achievable rate performance of the detectors under nonlinear distortion are shown in Fig. \ref{fig_nonlinear}.} For $\rm E_b/N_0$ from 5 dB to 30 dB, the DNN detector still can not overcome the overfitting and the performance of the \textcolor{black}{CycleDNN} detector also degrades significantly since the relationship between signals become more complex and more difficult to be captured. \textcolor{black}{Meanwhile, the proposed \textcolor{black}{CycleGAN} detector retains its superiority and the gap between the \textcolor{black}{CycleGAN} and the non-blind LMMSE detector is further expanded than in \mbox{section \ref{exp1}}. Note that for scenarios under low $\rm E_b/N_0$, the superiority of the \textcolor{black}{CycleGAN} detector over the non-blind LMMSE detector is insignificant since the risk of overfitting is high for DL models due to the effect of noise.} But as the $\rm E_b/N_0$ increases, the effect of noise is decreased so that the deep relationship between signals could be better captured, leading to better performance of the \textcolor{black}{CycleGAN} detector. \textcolor{black}{The decrease in pilot length also leads to worse detection performance but the training still remains stable.} Moreover, since neither the prior knowledge of the underlying channel model nor the CSI estimation for pre-decoding are required for the semi-blind \textcolor{black}{CycleGAN} detector, it indicates more practical value than the non-blind LMMSE detector.

\begin{figure}[!t]
\centering
\includegraphics[width=3.4in]{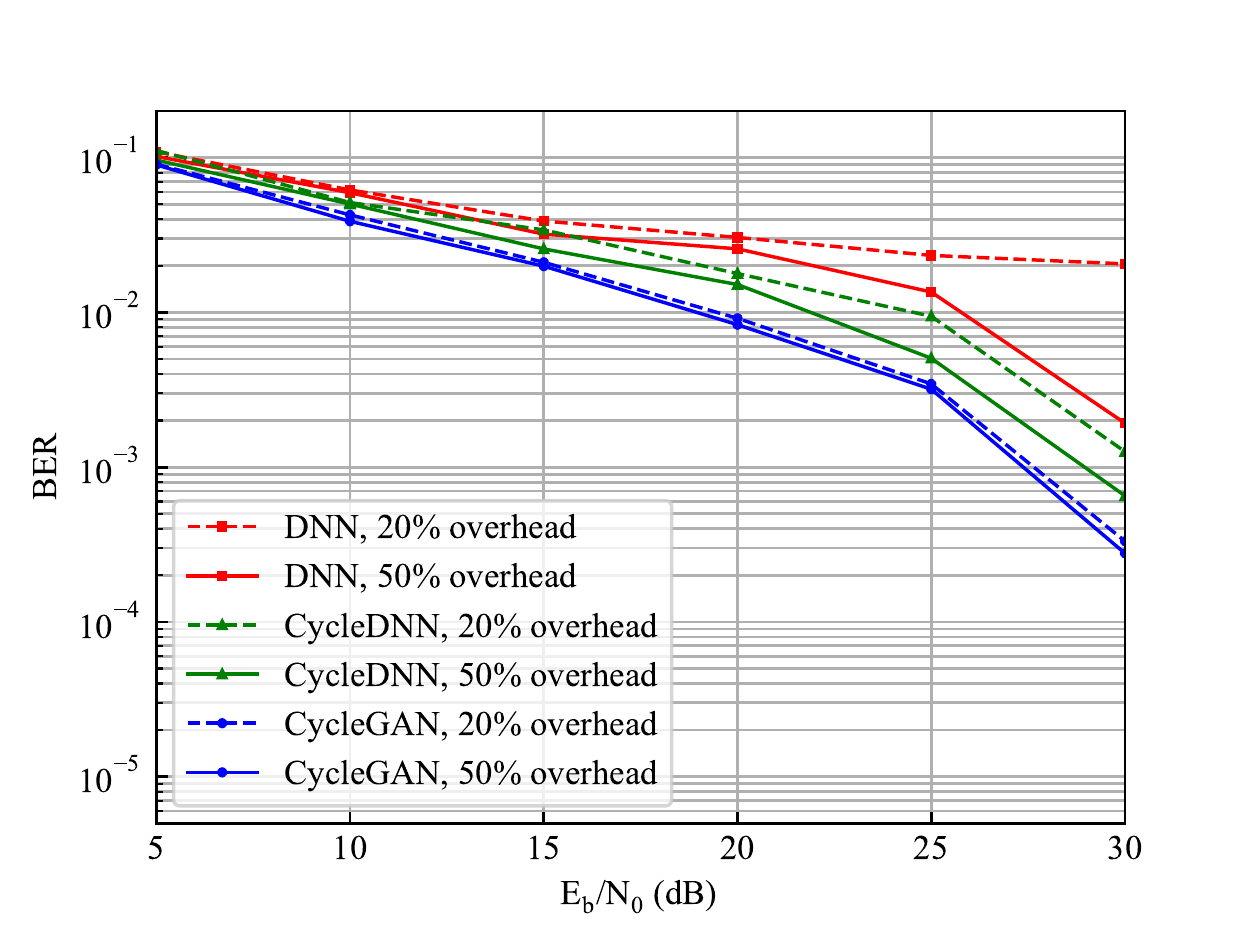}
\caption{\textcolor{black}{BER performance of the semi-blind \textcolor{black}{CycleGAN} detector, the semi-blind DNN detector, and the semi-blind \textcolor{black}{CycleDNN} detector with 20\% and 50\% overhead.}}
\label{fig_overhead}
\end{figure}

\subsection{Overhead Reduction \label{exp3}}
In this section we evaluate the benefit of the \textcolor{black}{CycleGAN} detector in reducing overhead. Since the experiments in above sections show that when the overhead is $20\%$, the DNN detector falls into severe overfitting and the \textcolor{black}{CycleDNN} detector can not accurately model the transmission process, in this \textcolor{black}{section,} we set the overhead to $50\%$ for the three DL detectors, i.e., increase the number of pilots to 160 symbols, and then compare the BER performance of them against those with 20\% overhead. Other simulation configurations are the same as in \textcolor{black}{Section} \ref{exp2}.

The simulation results are shown in Fig. \ref{fig_overhead}. Since the number of pilots is increased, the DNN detector and the \textcolor{black}{CycleDNN} detector can obviously better mitigate the overfitting and achieve quite lower BER than \textcolor{black}{those} with 20\% overhead. However, these two detectors still cannot outperform the \textcolor{black}{CycleGAN} detector with 20\% overhead even with increased training data. Note that for the \textcolor{black}{CycleGAN} detector, the increase of overhead does not bring significant improvement since 20\% overhead is already enough for the supervised training phase and the model can be further updated unsupervisedly by the received payload data even the corresponding transmitted signals are unknown. The results indicate that the \textcolor{black}{CycleGAN} detector can save over half of the requirement for overhead than standard DL models for similar BER performance. Such benefit in reducing the overhead is mainly achieved by virtue of the guidance from discriminators and the introduction of unsupervised learning. The guidance of discriminators can help the model to better capture the deep effects in the transmission process, such as the precoding and nonlinear distortion, while the introduction of unsupervised learning can make use of the received payload data to equivalently augment the training dataset. 

\begin{figure}[!t]
\centering
\includegraphics[width=3.4in]{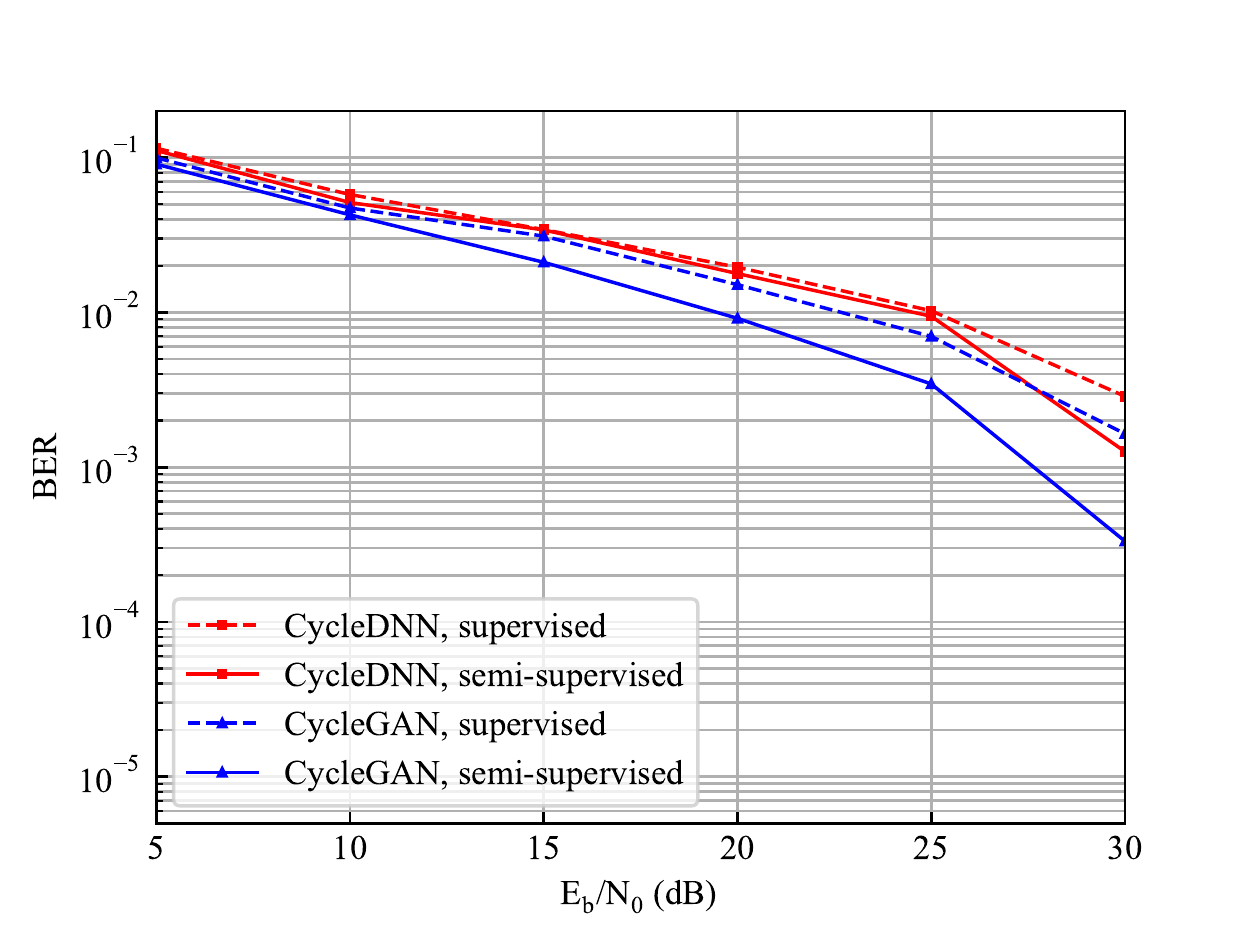}
\caption{\textcolor{black}{BER performance of the semi-supervised \textcolor{black}{CycleGAN} detector, the supervised \textcolor{black}{CycleGAN} detector, the semi-supervised \textcolor{black}{CycleDNN} detector and the supervised \textcolor{black}{CycleDNN} detector.}}
\label{fig_semi}
\end{figure}

\begin{figure*}[t]
\centering
\subfloat[BER under different channels.]{
    \centering
    \begin{minipage}[t]{3.4in}
        \centering       
        \includegraphics[width=\linewidth]{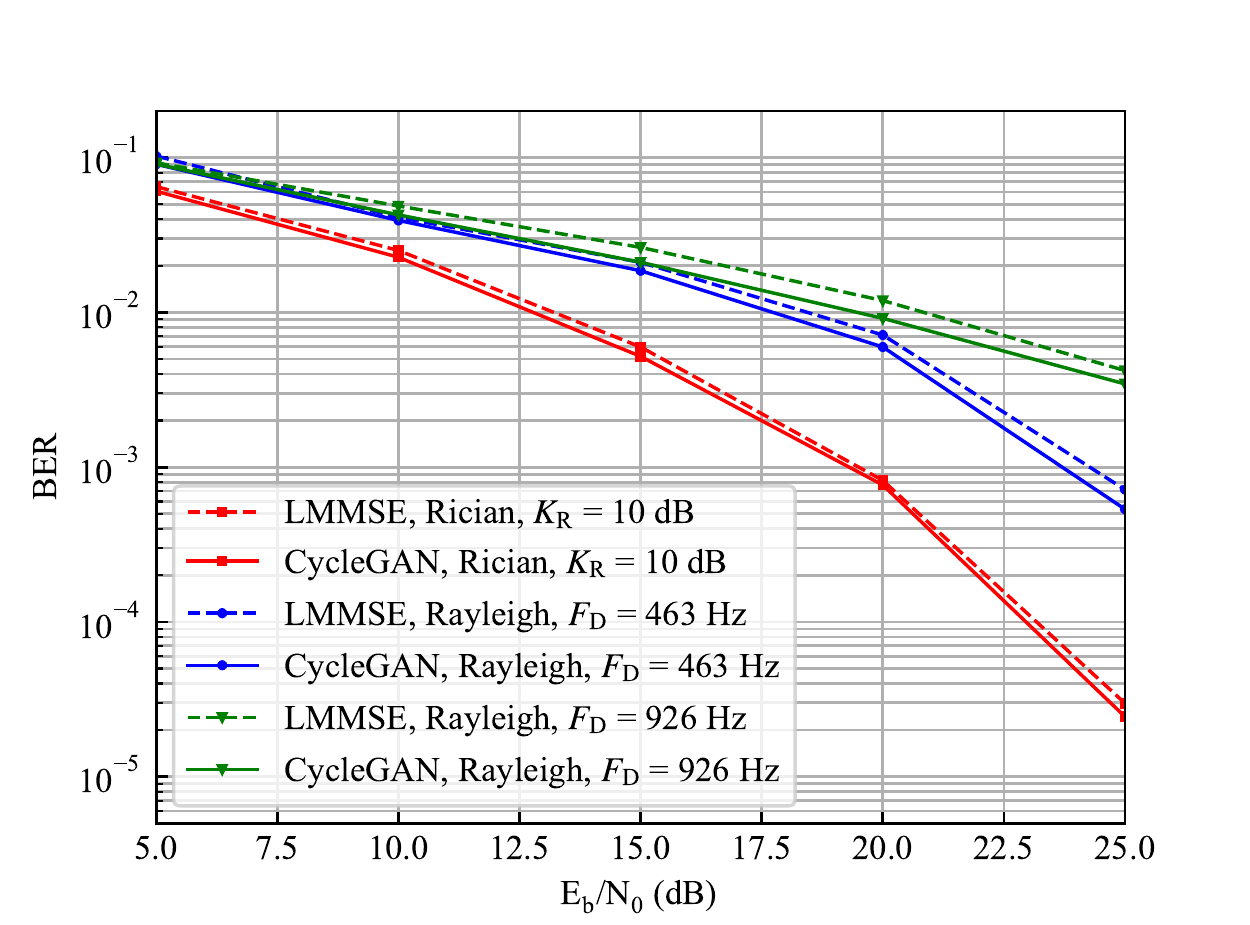}
        \end{minipage}
}%
\subfloat[Achievable rate under different channels.]{
    \centering
    \begin{minipage}[t]{3.4in}
        \centering       
        \includegraphics[width=\linewidth]{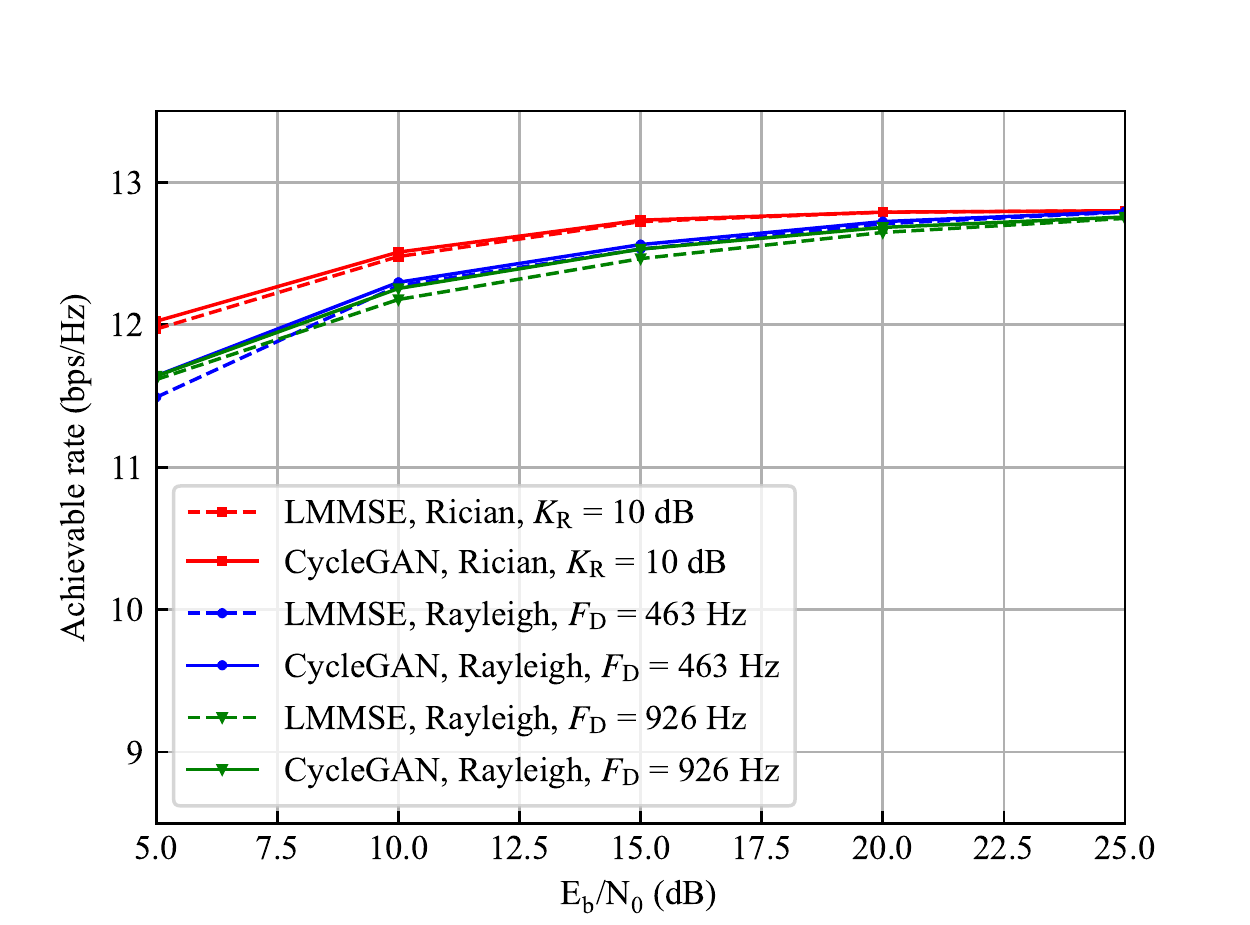}
    \end{minipage}
}%
\caption{\textcolor{black}{(a) BER and (b) achievable rate performance of the proposed \textcolor{black}{CycleGAN} and LMMSE under Rician fading channels with Rician factor of 10 dB, Rayleigh fading channels with maximum Doppler shift of 926 Hz, and Rayleigh fading channels with maximum Doppler shift of 463 Hz.}}
\label{fig_channels}
\end{figure*}

\subsection{Benefit of Semi-supervised Learning \label{exp4}}
To further evaluate the benefit of semi-supervised learning, we also investigate the BER performance of the \textcolor{black}{CycleGAN} detector trained by semi-supervised learning and the same model trained by supervised learning with pilots only, with other simulation conditions unchanged as in \textcolor{black}{Section} \ref{exp2}. Same comparison between the semi-supervised and supervised \textcolor{black}{CycleDNN} detector is also investigated for more general conclusion.    

Fig. \ref{fig_semi} shows that for $\rm E_b/N_0$ from 5 dB to 30 dB, the BER performance of the semi-supervised \textcolor{black}{CycleGAN} detector stays better than the supervised \textcolor{black}{CycleGAN} detector with over 10\% less BER, and the semi-supervised \textcolor{black}{CycleDNN} detector also outperforms the supervised \textcolor{black}{CycleDNN} detector especially for high $\rm E_b/N_0$. Note that the benefit of semi-supervised leaning for the \textcolor{black}{CycleDNN} detector is relatively limited since there is no guidance from discriminators so that it can be more difficult to avoid mislead of error bits in the unsupervised learning. Both the \textcolor{black}{CycleGAN} detector and the \textcolor{black}{CycleDNN} detector show the benefit of semi-supervised learning since plenty of new knowledge is provided by the received payload data during unsupervised learning, by which the model is able to better capture the distortion and other deep effects in the transmission process. Actually, the performance of the supervised \textcolor{black}{CycleGAN} detector is even worse than the non-blind LMMSE detector since the limited pilots do cause serious overfitting, and the semi-supervised learning makes great contributions to overcome this drawback.           

\subsection{\textcolor{black}{Detection Performance under Different Channels} \label{exp5}}
\textcolor{black}{In this section we investigate the BER and achievable rate performance of the proposed \textcolor{black}{CycleGAN} detector under different channel distributions. With other simulation conditions unchanged as in \ref{exp2}, we compare the performance of the proposed \textcolor{black}{CycleGAN} with non-blind LMMSE under Rician fading channel with Rician factor $K_{\rm R}$ of 10 dB, and Rayleigh fading channel with maximum Doppler shift $F_{\rm D}$ of 926 Hz and 463 Hz, respectively.}     
    
\textcolor{black}{Fig. \ref{fig_channels} shows that for $\rm E_b/N_0$ from 5 dB to 25 dB, the BER and achievable rate performance of the proposed \textcolor{black}{CycleGAN} detector retains its superiority under both Rician fading channels and Rayleigh fading channels with different setups of Doppler shifts. For Rician fading channels with Line-of-Sight \textcolor{black}{(LoS)} component, the proposed \textcolor{black}{CycleGAN} achieves better performance than under the Rayleigh fading channels. For the channels with larger Doppler shifts, degradation of the detection performance is deserved due to \textcolor{black}{severer} fading. Nevertheless, the training remains stable and the proposed \textcolor{black}{CycleGAN} detector still outperforms the non-blind LMMSE \textcolor{black}{detector}.} 

\section{Conclusion}
A semi-blind MIMO detection method based on \textcolor{black}{CycleGAN} is proposed, \textcolor{black}{which incorporates two modified LS-GANs into a bidirectional loop to improve the performance.} We trained the model under a semi-supervised learning strategy to make sufficient use of both the pilots and the received payload data. \textcolor{black}{Numerical results show that the BER and achievable rate performance of the proposed \textcolor{black}{CycleGAN} detector has a great superior over other benchmarks. The training of the proposed \textcolor{black}{CycleGAN} remains stable with different channel variation timescale and the semi-supervised learning makes the model a considerable progress. Another motivation of the proposed \textcolor{black}{CycleGAN} detector is that it requires quite few prior knowledge and assumptions, indicating reduced overhead and more promising application prospect for practical use.} %
\textcolor{black}{{\appendices
\section{Simulation Details of Channel Generation \label{channel generation}}
The Jakes' Model generates a Rayleigh fading channel subject to a given Doppler spectrum by synthesizing the complex sinusoids \cite{ref27}. The output sequence of the Jakes' Model is represented by 
\begin{equation}
\label{Jakes Model}
\mathbf{H}_{\rm{J}} = \{{h_k} \vert {h_k} \! =\! \frac{E_0}{\sqrt{2N_0 + 1}}[h_{\rm{I}}(t_{0} + k \times T_{\rm s}) + {\rm{j}}h_{\rm{Q}}(t_{0} + k \times T_{\rm s})]\} \text{,}
\end{equation}
where $k=1,2,3,...,K_{\rm s}$ is the index of samples, $K_{\rm s}$ is the number of samples, $E_0$ is the channel power, $N_0$ is a constant number of complex oscillators, $t_0$ is the initial time, $T_{\rm s}$ is the sampling time and ${\rm j}$ is the imaginary unit. $h_{\rm{I}}$ and $h_{\rm{Q}}$ are the real and imaginary parts, respectively, represented by 
\begin{equation}
\label{real part}
h_{\rm{I}}(t) = 2\sum_{n=1}^{N_0}({\rm{cos}}\phi_{n}{\rm{cos}}{\omega_{n}t}) + \sqrt{2}{\rm{cos}}\phi_{N}{\rm{cos}}\omega_{\rm{d}}t \text{,}
\end{equation}
and
\begin{equation}
\label{imaginary part}
h_{\rm{Q}}(t) = 2\sum_{n=1}^{N_0}({\rm{sin}}\phi_{n}{\rm{cos}}{\omega_{n}t}) + \sqrt{2}{\rm{sin}}\phi_{N}{\rm{cos}}\omega_{\rm{d}}t \text{,}
\end{equation}
where $\phi_{n}$ and $\phi_{N}$ are the initial phases of the $n$th Doppler shifted sinusoid and the maximum Doppler frequency, respectively, $\omega_{n}$ and $\omega_{\rm{d}}$ are the frequencies of the $n$th Doppler shifted sinusoid and the maximum Doppler frequency, respectively.}} 

\textcolor{black}{To generate a sequence of Rayleigh fading MIMO channel matrices, we use the Jakes' Model to generate each path of the channel with a random initial time. Then the output of our channel generator is represented by
\begin{equation}
\begin{split}
\label{Jakes Model 4 MIMO}
\mathbf{H} = &\{{h_{i,j,k}} \vert {h_{i,j,k}} = \frac{E_0}{\sqrt{2N_0 + 1}}[h_{\rm{I}}(t_{0}^{i,j}+ k \times T_{\rm s})\\ &+ {\rm{j}}h_{\rm{Q}}(t_{0}^{i,j} + k \times T_{\rm s})]\} \text{,}
\end{split}
\end{equation}
where $t_{0}^{i,j}$ is the initial time for the Jakes' Model of the path between the $j$th transmit antenna and the $i$th receive antenna.}

\textcolor{black}{From the perspective of time, each path of $\mathbf{H}$ is a continuous Rayleigh fading channel sequence generated by the Jakes' Model. While from the perspective of space, each sample of $\mathbf{H}$ for a given $k$ is equally a matrix of $i \times j$ random samples of the Jakes' Model among all time since the time term varies with $t_{0}^{i,j}$. Therefore, the entire $\mathbf{H}$ is approximately a sequence of continuous Rayleigh fading MIMO channel matrices, which is used for our simulations.}

\vfill


\begin{thebibliography}{1}

\bibitem{ref1}
H. Q. Ngo, A. Ashikhmin, H. Yang, E. G. Larsson, and T. L. Marzetta, ``Cell-free massive MIMO versus small cells,'' \textit{IEEE Trans. Wireless Commun.}, vol. 16, no. 3, pp. 1834--1850, Mar. 2017.

\bibitem{ref35}
W. Xu, Z. Yang, D. W. K. Ng, M. Levorato, Y. C. Eldar, and M. Debbah, ``Edge learning for B5G networks with distributed signal processing: semantic communication, edge computing, and wireless sensing," \textit{IEEE J. Sel. Topics Signal Process.}, vol. 17, no. 1, pp. 9--39, Jan. 2023.

\bibitem{ref2}
S. Buzzi and C. D’Andrea, ``Cell-free massive MIMO: User-centric approach,'' \textit{IEEE Wireless Commun. Lett.}, vol. 6, no. 6, pp. 706--709, Dec. 2017.

\bibitem{ref36}
W. Xu et al., ``Toward ubiquitous and intelligent 6G networks: from architecture to technology,” Sci. China Inf. Sci., vol. 66, no. 3, pp. 130300:1--2, Mar. 2023.

\bibitem{ref3}
N. Samuel, T. Diskin, and A. Wiesel, ``Learning to detect,'' \textit{IEEE Trans. Signal Process.}, vol. 67, no. 10, pp. 2554--2564, May. 2019.

\bibitem{ref4}
M. Zamanipour, ``A survey on deep-learning based techniques for modeling and estimation of massive MIMO channels,'' \textit{arXiv preprint arXiv:1909.05148}, 2019.

\bibitem{ref28}
N. Farsad and A. Goldsmith, ``Neural network detection of data sequences in communication systems,'' \textit{IEEE Trans. Signal Process.}, vol. 66, no. 21, pp. 5663--5678, Nov. 2018.

\bibitem{ref5}
M. Ghosh and C. L. Weber, ``Maximum-likelihood blind equalization,'' \textit{Opt. Eng.}, vol. 31, no. 6, pp. 1224--1229, Jun. 1992.

\bibitem{ref6}
L. Tong and S. Perreau, ``Multichannel blind identification: From subspace to maximum likelihood methods,'' \textit{Proc. IEEE}, vol. 86, no. 10, \mbox{pp. 1951--1968}, Oct. 1998.

\bibitem{ref7}
H. A. Cirpan and M. K. Tsatsanis, ``Maximum likelihood blind channel estimation in the presence of doppler shifts,'' \textit{IEEE Trans. Signal Process.}, vol. 47, no. 6, \mbox{pp. 1559--1569}, Jun. 1999.

\bibitem{ref8}
Y. Hama and H. Ochiai, ``Performance analysis of matched filter detector for MIMO systems in Rayleigh fading channels,'' in \textit{Proc. IEEE Global Commun. Conf.}, Singapore, Dec. 2017, pp. 1--6.

\bibitem{ref9}
A. Trimeche, N. Boukid, A. Sakly, and A. Mtibaa, ``Performance analysis of ZF and MMSE equalizers for MIMO systems,'' in \textit{Proc. Int. Conf. Design Technol. Integr. Syst.}, Tunis, Tunisia, May 2012, pp. 1--6.

\bibitem{ref10}
N. Kim, Y. Lee, and H. Park, ``Performance analysis of MIMO system with linear MMSE receiver,'' \textit{IEEE Trans. Wireless Commun.}, vol. 7, \mbox{no. 11}, pp. 4474--4478, Nov. 2008.
 
\bibitem{ref11}
Y. Gong, X. Hong, and K. F. Abu-Salim, ``Adaptive MMSE equalizer with optimum tap-length and decision delay,'' in \textit{Proc. Sensor Signal Process. Defence (SSPD)}, London, United Kingdom, Sep. 2010, pp. 1--5. 

\bibitem{ref12}
S. Chen, S. X. Ng, E. F. Khalaf, A. Morfeq, and N. D. Alotaibi, ``Multiuser detection for nonlinear MIMO uplink,'' \textit{IEEE Trans. Commun.}, vol. 68, no. 1, pp. 207--219, Jan. 2020.

\bibitem{ref34}
H. Huo, J. Xu, G. Su, W. Xu, and N. Wang, ``Intelligent MIMO detection using meta learning,'' \textit{IEEE Wireless Commun. Lett.}, vol. 11, no. 10, \mbox{pp. 2205}--2209, Oct. 2022.

\bibitem{ref13}
N. Samuel, T. Diskin, and A. Wiesel, ``Deep MIMO detection,'' in \textit{Proc. IEEE Int. Workshop Signal Process. Advances Wireless Commun.}, Sapporo, Japan, Jul. 2017, pp. 1--5.

\bibitem{ref14}
Q. Chen, S. Zhang, S. Xu, and S. Cao, ``Efficient MIMO detection with imperfect channel knowledge - A deep learning approach,'' in \textit{Proc. IEEE Wireless Commun. Netw. Conf. (WCNC)}, Marrakech, Morocco, Apr. 2019, pp. 1--6

\bibitem{ref15}
A. Caciularu and D. Burshtein, ``Blind channel equalization using variational autoencoders,'' in \textit{Proc. IEEE Int. Conf. Commun.}, Kansas City, MO, USA, May. 2018, pp. 1--6.

\bibitem{ref16}
A. Caciularu and D. Burshtein, ``Unsupervised linear and nonlinear channel equalization and decoding using variational autoencoders,'' \textit{IEEE Trans. Cogn. Commun. Netw.}, vol. 6, no. 3, pp. 1003--1018, \mbox{Sep. 2020}.

\bibitem{ref17}
X. Yi and C. Zhong, ``Deep learning for joint channel estimation and signal detection in OFDM systems,''  \textit{IEEE Commun. Lett.}, vol. 24, \mbox{no. 12}, pp. 2780--2784, Dec. 2020.

\bibitem{ref18}
E. Balevi, A. Doshi, A. Jalal, A. Dimakis, and J. G. Andrews, ``High dimensional channel estimation using deep generative networks,'' \textit{IEEE J. Sel. Areas Commun.}, vol. 39, no. 1, pp. 18--30, Jan. 2021.

\bibitem{ref33}
M. Vu and A. Paulraj, ``MIMO wireless linear precoding,'' \textit{IEEE Signal Process. Mag.}, vol. 24, no. 6, pp. 86--105, Sep. 2007.


\bibitem{ref19}
W. F. Al-Azzo, B. M. Ali, S. Khatun, N. K. Noordin, and S. M. Bilfagih, ``Peak-to-average power ratio reduction in OFDM systems using smoothing technique,'' in \textit{Proc. IEEE Malaysia Int. Conf. Commun.}, KualaLumpur, Malaysia, Dec. 2009, pp. 11--14.

\bibitem{ref20}
Y. Jeon, H. Do, S. Hong, and N. Lee, ``Soft-output detection methods for sparse millimeter-wave MIMO systems with low-precision ADCs,'' \textit{IEEE Trans. Commun.}, vol. 67, no. 4, pp. 2822--2836, Apr. 2019.

\bibitem{ref21}
Y. Suzuki and S. Narahashi, ``Power amplifier configuration for massive-MIMO transmitter,'' in \textit{Proc. Asia-Pacific Conf. Commun.}, Yogyakarta, Indonesia, Aug. 2016, pp. 38--43.

\bibitem{ref22}
X. Mao, Q. Li, H. Xie, R. Y. Lau, Z. Wang, and S. P. Smolley, ``Least squares generative adversarial networks,'' in \textit{Proc. IEEE Int. Conf. Comput. Vis. (ICCV)}, Venice, Italy, Oct. 2017, pp. 2794--2802.

\bibitem{ref32}
S. Nowozin, B. Cseke, and R. Tomioka, ``f-GAN: Training generative neural samplers using variational divergence minimization,'' in \textit{Proc. Adv. Neural Inf. Process. Syst.}, Barcelona, Spain, Dec. 2016, pp. 271--279.

\bibitem{ref30}
P. Isola, J. Zhu, T. Zhou, and A. A. Efros, ``Image-to-image translation with conditional adversarial networks,'' in \textit{Proc. IEEE Conf. Comput. Vis. Pattern Recognit. (CVPR)}, Honolulu, Hawaii, Jul. 2017, pp. 5967--5976.

\bibitem{ref31}
M. Mirza and S. Osindero, ``Conditional generative adversarial nets,'' \textit{arXiv preprint arXiv:1411.1784}, 2014.

\bibitem{ref23}
S. Iizuka, E. Simo-Serra, and H. Ishikawa, ``Globally and locally consistent image completion,'' \textit{ACM Trans. Graph.}, vol. 36, no. 4, \mbox{pp. 1--14}, Aug. 2017.

\bibitem{ref24}
J. Zhu, T. Park, P. Isola, and A. A. Efros, ``Unpaired image-to-image translation using cycle-consistent adversarial networks,'' in \textit{Proc. IEEE Int. Conf. Comput. Vis. (ICCV)}, Venice, Italy, Oct. 2017, pp. 2242--2251.

\bibitem{ref25}
G. E. Hinton, N. Srivastava, A. Krizhevsky, I. Sutskever, and R. R. Salakhutdinov, ``Improving neural networks by preventing co-adaptation of feature detectors,'' \textit{arXiv preprint arXiv:1207.0580}, 2012.

\bibitem{ref26}
N. Srivastava, G. E. Hinton, A. Krizhevsky, I. Sutskever, and R. R. Salakhutdinov, ``Dropout: A simple way to prevent neural networks from overfitting,'' \textit{J. Mach. Learn. Res.}, vol. 15, no. 1, pp. 1929--1958, \mbox{Jan. 2014}.

\bibitem{ref27}
W. C. Jakes, ``Multipath Interference," \textit{Microwave Mobile Communications}, D. C. Cox, Ed. New York: Wiley-IEEE, 1974, Ch. 1, pp. 11--78.
\bibitem{ref29}
Y. Bengio, ``Learning deep architectures for AI,'' \textit{Found. Trends Mach. Learn.}, vol. 2, no. 1, pp. 1--127, Jan. 2009.



\end{thebibliography}
\end{document}